\documentclass[11pt]{article}

\pdfoutput=1

\usepackage[margin=1in]{geometry}
\usepackage{amsmath}
\usepackage{graphicx}
\usepackage{enumerate}
\usepackage{natbib}
\usepackage{url} 

\usepackage[T1]{fontenc}
\usepackage[latin9]{inputenc}
\usepackage{array}
\usepackage{longtable}
\usepackage{float}

\usepackage{amsthm}
\usepackage{mathtools}
\usepackage{amssymb}

\usepackage{setspace}
\usepackage{esint}

\usepackage{dsfont}
\usepackage{amsfonts}
\usepackage{csvsimple}
\usepackage{fixltx2e} 

\usepackage{subcaption}
\usepackage{booktabs}
\usepackage[dvipsnames]{xcolor}

\usepackage{caption}
\usepackage{babel}
\usepackage{authblk}
\usepackage{algorithm}
\usepackage[noend]{algpseudocode}

\usepackage{csvsimple}
\usepackage{hyperref}

\newtheorem{theorem}{Theorem}[section]

\newtheorem{prop}[theorem]{Proposition}

\newcommand{\blind}{1}

\usepackage{hyperref}

\begin{document}
	
	\def\spacingset#1{\renewcommand{\baselinestretch}%
		{#1}\small\normalsize} \spacingset{1}

	
	\if1\blind
	{
		\title{\bf Joint Modeling for Geometry and Functionality of Cerebral Cortical Surface Images}
		
		\author[1]{Jingjing Zou \thanks{The authors gratefully acknowledge the help and comments of Professor Lexin Li during the time this work was carried out. JA's work was partially supported  by the Engineering and Physical Sciences Research Council (EP/T017961/1).} }
		
		\author[2]{Chi-Hua Chen}
		
		\author[3]{John A. D. Aston}
		
		\affil[1]{Herbert Wertheim School of Public Health and Human Longevity Science, University of California, San Diego}
		
		\affil[2]{Department of Radiology, University of California, San Diego}
		
		\affil[3]{Statistical Laboratory, DPMMS, University of Cambridge}	
		\maketitle		
	} \fi
	
	\if0\blind
	{
		\bigskip
		\bigskip
		\bigskip
		\begin{center}
			{\LARGE\bf Joint Modeling for Geometry and Functionality of Cerebral Cortical Surface Images}
		\end{center}
		\medskip
	} \fi

	\begin{abstract}
		We propose a framework for jointly modeling the geometry and functionality in high dimensional functional surfaces. The proposed mixed effects model characterizes effects of subject-specific covariates and exogenous stimuli on functional surfaces while accounting for potential mutual-influence of their geometry and functionality. This is achieved through a computationally efficient estimation method that incorporates regularized estimation of the precision matrix of the random effects. We perform a thorough analysis of cerebral cortical surface structural MRI and task fMRI data from the Human Connectome Project and discover relationships between the geometric shapes of cortical surface and neuronal activation responding to task stimuli. Our findings highlight new modes of correspondence between cortical surface shape and functional activation relevant to emotion processing.
	\end{abstract}
	
	\noindent%

	\spacingset{1.9} 

	\section{Introduction}
	
	Previous studies have shown that the structure and function of the human brain are interconnected. 
	For example, 
	\cite{Stahn:2019} discovered that after a 14-month journey in Antarctica, nine expeditioners lost significant volume in the hippocampus, a brain component critical for learning, memory and emotion processing, due to prolonged social isolation.
	
	Multiple types of magnetic resonance imaging (MRI) have served as important tools for studying the brain \citep{Huettel:2009uc}.
	The structural MRI measures aspects of the anatomy of the brain. 
	The functional MRI (fMRI) examines brain functions via the blood-oxygen-level dependent (BOLD) signals that measure the underlying neuronal activities.
	In particular, the cortical surface fMRI, in which the BOLD signals are mapped to a two-dimensional (2D) manifold representing the cerebral cortex, has attracted increasing attention in recent studies. 
	Comparing to traditional 3D volumetric MRI, it provides a more intuitive representation of functional distance (\cite{Glasser:2013gd, Mejia:2017uh}).	
	
	The integration of cortical surface structural and functional MRI provides a comprehensive view of both the structure and function of the brain, allowing for the examination of the relationship between the brain's geometric shape and its functionality, in addition to individual analyses of each component.
	However, the relationship between the geometry and functionality of the brain is complex and not fully understood. In order to capture this relationship, it is important to study both components of the brain in a holistic manner.	To this end, we propose a novel statistical model framework for jointly modeling the geometry and functionality of the brain, as represented by vertex-level cortical surface structural and task functional MRI. The proposed model simultaneously captures effects of covariates such as environmental, demographic and social variables as well as exogenous stimuli on the variability in the geometry and function of the brain and potential correspondence between the geometry and functionality.

	\subsection{Past Work on Analyzing Cortical Surface Images}
	
	Past work on analyzing cortical surface imaging data typically focused either on the structure of the cortical surface with summarizing features (\cite{Im2008}), or on the functionality characterized by the fMRI.
	For analysis of the fMRI, a widely used approach is to fit separate generalized linear models (GLM) at individual vertices on the cortical surface to assess the relationship between functional activity and relevant covariates.
	Adjustment for multiple comparison is needed and achieved by thresholding or clustering methods.
	Examples of studies using this approach include
	\cite{Friston:1994kt, Worsley:1995vs, Worsley:1996gj, Andrade:2001ec, Genovese:2002bx, HaglerJr:2006gi, Smith:2009gi, Lindquist:2015fx, Barch:2013kqb, Poline:2016gb}. 
	Studies have demonstrated potential pitfalls of this approach including reduced power (\cite{Ishwaran:2011ig}) and sensitivity to voxel/vertex size and smoothing (\cite{Woo:2014fj}, \cite{Mejia:2017uh}).
	
	An important issue with vertex-wise approaches is that they neglect the underlying spatial dependence among vertices and information contained in the totality of the data.
	Methods incorporating spatial dependence have been proposed to address this issue (\cite{Formisano:2004iy, Friston:2002fs, Friston:2002gc, Penny:2005do, Siden:2017cm, Mejia:2017uh}).
	In these studies, different geometric shapes of subjects were registered to a common template and functional signals were aligned according to the registration. However, analyzing the aligned functional signals alone eliminates information on geometric variability among the subjects and excludes information on associations between the geometry and functionality of cortical surfaces from subsequent analyses. 
	
	In recent studies aimed at simultaneously modeling the variability in the geometry and the functionality of the brain,
	\cite{Lila:2020} proposed an extension of the metamorphosis model (\cite{Charlier:2015fh}): random variables were incorporated into the metamorphoses representing the shape and functionality of the cerebral cortex to capture the geometric and functional variability. This approach, however, cannot be generalized to tasks/covariates-based studies without further development.
	\cite{lila2021functional} proposed a metamorphosis-based model to analyze the shape and functional connectivity of the cortical surface. The functional connectivity was measured by the second-order statistic of the covariance matrix among resting state fMRI BOLD signals at different vertices.
	This method is not suitable for analyzing the first-order representations of functionality, such as fMRI BOLD signals at different vertices or regions of interest. Additionally, the calculation of the covariance matrix involves compressing the temporal dimension, making it impossible to perform subsequent analysis involving any time-varying covariates effects.

	\subsection{The Proposed Scheme of Analysis}
	
	Despite previous studies, questions remain in studying the geometry and functionality of cortical surfaces, particularly in task based studies.
	The first question is how to model the variability in both geometry and functionality of the cortical surface and to what extent the variability can be explained by covariates, including demographics and health measurements, and exogenous stimuli in tasks.
	There are three challenges in this task:
	a). The functionality represented by fMRI BOLD signals is defined on distinct subject-specific domains of cortical surfaces and needs to be mapped to a common domain.
	b). The dimensionality of both the geometric shapes and functionality are large and requires dimension reduction methods.
	c). The functionality represented by the task fMRI signals are expected to vary in response to different phases of tasks, and modeling of the varying functions requires the preservation of the temporal dimension and accounting for potential temporal correlations in the functionality.
	
	The second question is how to capture correlations between the geometry and the functionality of the cortical surfaces to uncover meaningful connections between the shape and functions of the brain.
	In examining such correlations, effects of covariates including exogenous stimuli and potential confounders on the geometry and functionality need to be  disentangled from the variability in the geometry and functionality.
	Failure to do so may result in biased findings when evaluating the covariates' effects on the geometry and the functionality.
	Therefore, the effects of the covariates on the geometry and the functionality cannot be sufficiently analyzed in separate models, and a unified model to account for the covariates' effects as well as the correlations is required.

	To address these unsolved issues, we develop a unified mixed effects model suited for analyzing jointly the cortical surface structural and functional MRI including task fMRI.
	The multi-variate outcomes are geometric and functional PC scores, obtained using the stochastic metamorphism model (\cite{Lila:2020}) and a functional principal component analysis (fPCA).
	Effects of subject-specific covariates and exogenous stimuli are modeled with fixed effects while mutual-influences of the geometry and the functionality are modeled with random effects.
	The structure and positive-definiteness of the covariance of the random effects are guaranteed via the proposed parameterization of the Cholesky decomposition of the precision (inverse of the covariance) matrix.
	
	We propose a scalable iterative estimation method for the mixed effects model utilizing results in \cite{Bickel:2008eza} that the estimation of the Cholesky decomposition of the precision matrix can be converted to parallel regularized regressions. 
	Studies with similar focuses can be found in, for example, \cite{Levina:2008cn, Rothman:2008bi, Rocha:GZtp9s91} and the references therein.
	Comparing to the Restricted Maximum Likelihood Estimation (REML, c.f. \cite{ThompsonJr:1962gv, Jiang:1996if}) methods,
	the proposed approach is computationally efficient, especially for the high dimensionality of this particular application, and is more flexible in parameterizing the precision/covariance matrices while ensuring the positive definiteness.

	\subsection{Data Motivation: the Human Connectome Project (HCP)} \label{sec: HCP motivation}
	
	We conduct a comprehensive analysis on the cortical surface task fMRI data of 100 unrelated young healthy adults from the Human Connectome Project (HCP) (\cite{HCPmanual}), focusing on measures taken from an emotion processing task (\cite{Hariri:2002cl}).
	During the task, participants were asked to match pictures on the bottom and top of the screen, which showed either faces with emotions of anger or fear, or geometric shapes as the control.
	Tasks were presented to the participants in blocks, each consisted of six 3-second trials of the same type (face or shape).
	For each individual participant, 176 tfMRI scans were taken during the trials and the intermittent resting periods in-between.
	In addition to the task-fMRI scans, a one-time structural MRI was used to measure the $xyz$ coordinates of each participant's cortical surface.
	
	With the proposed approach, we reveal effects of covariates including age and gender on both the geometric shapes of the cortical surfaces and the activated functional regions identified by the fMRI, as well as effects of task stimuli on the functional activated regions.
	Age and gender are found to have more significant associations with the functionality than with the geometry of the cortical surface.
	Associations between the emotion processing task stimuli and activated regions indicate that functional areas associated with the emotion processing include the inferior and middle temporal, somatosensory and motor cortex, auditory association cortex, orbital and polar frontal cortex, primary visual cortex, para-hippocampal, and fusiform.
	Results also demonstrate the advantage of the proposed approach in statistical power when revealing meaningful associations between the covariates and the functionality of the cortical surface due to its ability of handling the task fMRI time series in their full temporal dimension.
	Furthermore, the proposed approach enables discovery of several modes of correspondence between geometric shapes and functionality of cortical surfaces in processing emotions. This gives concrete evidence that a joint approach to shape and functional information is required. Regions of activation in parts of the inferior temporal, occipitotemporal sulcus, fusiform gyri and temporal lobe are found to have potential correspondences with the geometric shapes of these regions.
	To our knowledge, this study is among the first to model the geometry and functionality of cortical surface structural MRI and task fMRI jointly. Given the emerging opportunities and challenges brought by the relatively new cortical surface data format, our study provides a novel tool for the new data format to address important substantive problems arisen in the study of the cerebral cortex.
	
	The paper is organized as follows.
	Section \ref{sec: reg model} introduces the statistical representation of cortical surface images and formulates the proposed mixed effects model.
	Section \ref{section: estimation} delineates the estimation pipeline.
	Section \ref{sec: HCP data} studies the HCP data with the proposed methods and compares findings with existing clinical studies. Some concluding remarks are then provided. In the Appendix, we set out some simulations studies which provide further justification for the methodology proposed and comparisons of the proposed approach to existing methods including REML.
	Additional technical details are provided in the Supplementary Material.

	\section{A Mixed Effects Model for Cortical Surfaces tfMRI} \label{sec: reg model}
	
	\subsection{The Functional Manifold Model} \label{sec: model for functional surf}
	
	Suppose there are $ N $ subjects, each has a time series of $T$ task fMRI scans and a structural MRI.
	The data of the $i$th subject are denoted by
	$\{\mathcal{M}_i, Y_i(1), \dots, Y_i(T) \}$,
	where
	$\mathcal{M}_i$ is a time-invariant two-dimensional manifold embedded in $\mathbb{R}^3$ denoting the geometric shape of cortical surface measured by the $xyz$ coordinates of the structural MRI, 
	and $\{Y_i(1), \dots, Y_i(T)\}$ are square integrable functions on $\mathcal{M}_i$ denoting the time series of functional fMRI blood-oxygen-level-dependent (BOLD) signals.
	
	
	We adopt the stochastic metamorphosis model in \cite{Lila:2020} to characterize the variability in the geometry and functionality of the cortical surfaces.
	Suppose
	$\mathcal{M}_{i} = \varphi(v_i, \cdot) \circ \mathcal{M}_{0}$,
	where $ \mathcal{M}_{0}$ is a template manifold, usually the mean shape of the subjects' cortical surfaces,
	and $ \varphi(v_i, \cdot): \mathbb{R}^3 \rightarrow \mathbb{R}^3 $ is a deformation operator indexed by a subject specific $v_i$,
	where $\{v_i: i =1,\dots, N\}$
	are independent and identically distributed (i.i.d.) samples of a zero-mean random field $ V $ defined in a Hilbert space.
	Subject to additional constraints,
	$\varphi(v_i, \cdot)$ can be further represented as a diffeomorphism
	determined by subject-specific \textit{initial momenta} $m^i_{0}$ defined on a finite set of control points on $ \mathcal{M}_{0}$.
	Intuitively, the initial momenta are vectors in $\mathbb{R}^3$ at selected vertices of the template manifold,
	indicating the directions and magnitudes to ``drag'' the template $\mathcal{M}_{0}$ towards the subject-specific shape $\mathcal{M}_{i}$.
	For technical details, see \cite{Miller:2006ft, Younes:2010eu, Charlier:2015fh, Lila:2020}.

	The functionality $Y_i(t)$ defined on the domain of $\mathcal{M}_{i}$ can be mapped to the domain of $\mathcal{M}_{0}$ with the
	inverse deformation $\varphi^{-1}(v_i, \cdot)$. 
	Suppose
	$Y_i(t) = X_i(t) \circ \varphi^{-1}(v_i, \cdot)$
	then $\{X_i(t): t = 1, \dots, T\}$ denote the fMRI signals mapped to the template.
	In some special cases, a correspondence map exists between the subject-specific manifold and the template manifold, and the functionality on the subject's manifold can be mapped directly to that on the template without using the inverse deformation.

	Principal component models are further assumed to decompose the variability in the geometry and functionality.
	Suppose
	$ X_i(t)  = \mu(t) + \delta Z_i(t)$, where $ \mu $ is the mean function, $ \delta $ is a scalar, and the $ \{Z_i(t): i = 1, \dots, N\} $ are realizations of a discrete time stochastic process $ \{Z(1),\dots, Z(T)\} $ where each $ Z(t) \in L^2(\mathcal{M}_0) $.
	Suppose a finite number of principal components are adequate for explaining most of the variability:
	$v_i = \sum_{j = 1}^{K^G} a_{ij}^G \psi_j^G$ and $Z_i(t) = \sum_{j = 1}^{K^F} a_{ij}^F(t) \psi_j^F$,
	where $ K^G $ and $ K^F $ are numbers of geometric and functional PCs, respectively.
	Here $ a^G_{ij} $ is the projection coefficient of $ v_i $ on the $ j $th geometric basis $\psi_j^G$, 
	and $ a^F_{ij}(t) $ is the projection coefficient of $ Z_i(t) $ on the $ j $th functional basis $\psi_j^F$ (\cite{Riesz:1955ul}).

	\subsection{The Mixed Effects Model}
	
	In this section we introduce the main model for effects of subject-specific covariates on the geometry and functionality of cortical surface as well as the correlation between the geometry and functionality.
	Let $ \mathbf{a} = (\mathbf{a}^G, \mathbf{a}^F)$ be the combination of geometric and functional PC projection coefficients (scores).
	Here
	$\mathbf{a}^G \in \mathbb{R}^{N \times K^G} $ with $a^G_{ij}$ denoting the $i$th subject's projection coefficient on the $j$th geometric PC.
	$\mathbf{a}^F \in \mathbb{R}^{N \times (T \cdot K^F)} $ with each row
	$ \mathbf{a}_i^F = (\mathbf{a}_{i1}^F, \dots, \mathbf{a}_{i K^F}^F) $ 
	and $ \mathbf{a}_{ik}^F = ({a}_{ik}^F(1), \dots, {a}_{i k}^F(T)) $ denotes the $ i $th subject's projection coefficients on the $k$th functional principal component at time $t= 1,\dots, T$.
	With slight abuse of notation, a simplified version of the proposed mixed effect models can be written as
	\begin{align*} 
		\mathbf{a}^G & =  \mathbf{U} \boldsymbol{\alpha}^G +  \boldsymbol{\gamma}^G + \boldsymbol{\epsilon}^G,  \\
		\mathbf{a}^F & = \mathbf{U} \boldsymbol{\alpha}^G +  \mathbf{W} \boldsymbol{\beta} + \boldsymbol{\gamma}^F + \boldsymbol{\epsilon}^F.
	\end{align*}
	This model characterizes the covariates' effects on the geometry and functionality with fixed effects $\boldsymbol{\alpha}^G$, $\boldsymbol{\alpha}^F$, and $\boldsymbol{\beta}$ while accounting for the correlations between the geometry and functionality with random effects $\boldsymbol{\gamma}^G$ and $\boldsymbol{\gamma}^F$.
	In what follows, we describe in detail the parameterizations and re-shaping of each component that guarantee the correct structures and dimensionalities of the model.
	
	The fixed effects of two types of covariates are of interest. The first type is time-invariant covariates such as age and gender which have constant values during  the tasks.
	We denote the number of time-invariant covariates by $ p_u $ and
	let $ \textbf{U} = (\textbf{U}_1, \dots, \textbf{U}_{p_u}) \in \mathbb{R}^{N \times p_u}$ be the matrix of all time-invariant covariates.
	The second type is time varying covariates such as the task signals/phases in studying the brain's response to exogenous stimuli using task fMRI.
	Denote the number of time varying covariates by $p_w$.
	Let $ \mathbf{W} = (\textbf{W}_{1}, \dots, \textbf{W}_{p_w}) \in \mathbb{R}^{(NT) \times p_w} $,
	where $ \textbf{W}_{k} = (W_{1,k}(1), \dots, W_{N,k}(1), \dots, W_{1,k}(T), \dots, W_{N,k}(T))^\intercal$ denotes the column vector of the time series of the $k$th time-varying covariate for all $N$ subjects.
	We assume the functionality represented by $ \{\mathbf{a}_i^F \} $ is affected by both $ \mathbf{U} $ and $ \mathbf{W} $,
	while $ \{\mathbf{a}_i^G \}$ is affected by $\mathbf{U}$ only.
	This is because the geometric shape of a subject is considered to be the same during the short period of task fMRI scanning.
	This could be relaxed in the case of longitudinal imaging studies.
	Let $ \boldsymbol{\alpha}^G  $ denote effects of time-invariant covariates $ \mathbf{U} $ on the geometric projections $ \mathbf{a}^G $,
	$ \boldsymbol{\alpha}^F $ denote the effect of $ \mathbf{U} $ on functional projection coefficients $ \mathbf{a}^F $,
	and $ \boldsymbol{\beta} $ denote the effect of the time-varying covariates $\mathbf{W}$ on $ \mathbf{a}^F $.

	
	To model the correlations between the geometry and functionality, we introduce random effects $ \boldsymbol{\gamma} = (\boldsymbol{\gamma}^G, \boldsymbol{\gamma}^F) $, where  $ \boldsymbol{\gamma}^G \in \mathbb{R}^{N \times K^G} $ and $ \boldsymbol{\gamma}^F \in \mathbb{R}^{N \times (T \cdot K^F)} $.
	Assume the rows  $\boldsymbol{\gamma}_i \stackrel{\text{iid}}{\sim} N(0, \boldsymbol{\Sigma}_\gamma) $, where $\boldsymbol{\Sigma}_\gamma$ plays the important role in capturing the correlations between the geometry and functionality and is assumed to follow the structure:
	\begin{align} \label{eq: Sigma_gamma_tilde}
		&  \boldsymbol{\Sigma}_{\boldsymbol{\gamma}}
		=\begin{bmatrix}
			\boldsymbol{\Sigma}_{GG} & \boldsymbol{\Sigma}_{GF_1} & \cdots & \cdots & \boldsymbol{\Sigma}_{GF_{K^F}} \\
			\boldsymbol{\Sigma}_{GF_1}^\intercal & \boldsymbol{\Sigma}_{F_{1}F_{1}} & \boldsymbol{0} & \cdots & \boldsymbol{0} \\
			\vdots & \boldsymbol{0} & \ddots & \ddots & \vdots \\
			\vdots & \vdots & \ddots & \ddots & \boldsymbol{0} \\
			\boldsymbol{\Sigma}_{GF_{K^F}}^\intercal & \boldsymbol{0} &  \cdots & \boldsymbol{0} & \boldsymbol{\Sigma}_{F_{K^F}F_{K^F}}
		\end{bmatrix}.
	\end{align}
	In \eqref{eq: Sigma_gamma_tilde},
	$\boldsymbol{\Sigma}_{GG} = \text{diag} \{\sigma^2_{G_1}, \dots,  \sigma^2_{G_{K^G}} \} $ is the covariance matrix of $(\boldsymbol{\gamma}^G_{i, 1}, \dots,  \boldsymbol{\gamma}^G_{i, K^G})$ for any $i$.
	For each $i $ and $k = 1, \dots, K^F$,
	$\boldsymbol{\Sigma}_{F_{k} F_{k}}$ is the temporal covariance matrix among
	$\{\boldsymbol{\gamma}^F_{i, (k-1)T+t}: t = 1, \dots, T\}$.
	For each $j \in \{1, \dots, K^G\}$ and $t \in \{1, \dots, T\}$,
	$\boldsymbol{\Sigma}_{GF_{k}}(j, t)$ is the covariance between $\boldsymbol{\gamma}^G_{i, j}$ and $\boldsymbol{\gamma}^F_{i,(k-1)T+t}$.
	The off-diagonal elements in $\boldsymbol{\Sigma}_{GG}$ and the non-diagonal blocks except for $\{\boldsymbol{\Sigma}_{GF_{k}}, \boldsymbol{\Sigma}_{GF_{k}}^\intercal: k = 1, \dots, K^F \}$ are assumed to be zero, reflecting the zero correlations among the projection coefficients on the geometric PCs and among the projections on the functional PCs.

	Let $ \boldsymbol{\epsilon}^G \in \mathbb{R}^{N \times K^G} $ and $ \boldsymbol{\epsilon}^F \in \mathbb{R}^{N \times (T K^F)}$ denote the independent errors
	in measuring the geometry and functionality that cannot be explained by the fixed effects nor the random effects.
	They are assumed to be independent of $\boldsymbol{\gamma}$.
	Assume the rows $ \boldsymbol{\epsilon}^G_i \sim N(\boldsymbol{0}, \sigma_\epsilon^2 I_{K^G})$,
	and $ \boldsymbol{\epsilon}^F_i $ follows i.i.d.
	$ N(\boldsymbol{0}, \sigma_\epsilon^2 \mathbf{I}_{T \cdot K^F})$.
	Let $\text{vec}(\cdot)$ indicate the operator that reshapes a matrix to a vector by column, then
	the joint distribution of $ \mathbf{a}^G$ and $ \mathbf{a}^F  $ is characterized by the mixed effects model, now written in proper formats with dimensionalities of each component marked in the subscripts:
	\begin{align} \label{eq: mixed model}
		\mathbf{a}^G_{N \times K^G} & =  \mathbf{U}_{N \times p_u} \boldsymbol{\alpha}^G_{p_u \times K^G} +  \boldsymbol{\gamma}^G_{N \times K^G} + \boldsymbol{\epsilon}^G_{N \times K^G},  \\
		\text{vec}\big[\mathbf{a}^F_{N \times (T \cdot K^F)}\big]
		& = \text{vec} \big[(\mathbf{U}_T, \mathbf{W})_{(NT) \times (p_u + p_w)} \cdot
		(\boldsymbol{\alpha}^F_{K^F \times p_u}, \boldsymbol{\beta}_{K^F \times p_w})^\intercal \big]
		+\text{vec} \big[
		\boldsymbol{\gamma}^F_{N \times (T K^F)} + \boldsymbol{\epsilon}^F_{N \times (T K^F)} \big], \nonumber 
	\end{align}
	where
	$ \mathbf{U}_T = (\mathbf{U}^\intercal, \dots, \mathbf{U}^\intercal)^\intercal $ is the expansion of $ \mathbf{U} $ by repeating it $T$ times.
	
	
	
	%
	%

	\subsection{Parameterization of $\boldsymbol{\Sigma}$} \label{sec: spd}
	The covariance matrix of the outcome $(\mathbf{a}^G, \text{vec}[\mathbf{a}^F])$ in \eqref{eq: mixed model} is
	$\boldsymbol{\Sigma} = \boldsymbol{\Sigma}_{\boldsymbol{\gamma}} + \sigma_\epsilon^2 \mathbf{I}_{K^G + T \cdot K^F}$.
	Not all parametrizations of $\boldsymbol{\Sigma} $ lead to positive definite matrices required for a \textit{bona fide} covariance matrix,
	even if each of the blocks $\boldsymbol{\Sigma}_{GG}$, $\boldsymbol{\Sigma}_{F_{k}F_{k}} $ and $\boldsymbol{\Sigma}_{GF_k}$ in $\boldsymbol{\Sigma}_{\boldsymbol{\gamma}}$ is a well defined covariance matrix.
	To address this issue we parameterize the precision matrix $\boldsymbol{\Sigma}^{-1}$ with the following Cholesky decomposition
	\begin{equation} \label{eq: decomp_precision}
		\boldsymbol{\Sigma}^{-1} = \mathbf{L}^\intercal  \mathbf{D}^{-1} \mathbf{L},
	\end{equation}
	where $\mathbf{D}$ is a diagonal matrix with all positive elements and $\mathbf{L}$ is a lower triangular matrix with all diagonal elements equal to 1.
	Then the covariance matrix has the Cholesky decomposition
	$\boldsymbol{\Sigma} = {\mathbf{L}}^{-1} \mathbf{D} ({\mathbf{L}}^{-1})^\intercal$,
	where ${\mathbf{L}}^{-1}$ is also lower triangular, which guarantees the positive definiteness of $\boldsymbol{\Sigma}$.
	
	For a random vector $ (\gamma_1, \dots ,\gamma_p)$ from any distribution with mean $\boldsymbol{0}$ and covariance matrix $\boldsymbol{\Sigma}$, elements of $\mathbf{D}$ and $\mathbf{L}$ in \eqref{eq: decomp_precision} can be written as regression coefficients (\cite{pourahmadi99}, \cite{Levina:2008cn}, \cite{Rothman:2010cy}).
	For $j > 1$, let
	$\gamma_{j} = \sum_{k=1}^{j-1} \zeta_{jk} \gamma_{k} + \xi_{j}$,
	where $\{\zeta_{jk} \}$ are the coefficients of the best linear predictor of $\gamma_{j}$ by $\{\gamma_{k}: k = 1, \dots, j-1 \}$ and $\xi_{j}$ is the residual.
	Let $\xi_{1} = \gamma_{1}$, $ \sigma_j^2 = \text{Var}(\xi_{j})$ and $\boldsymbol{\zeta} = \{\zeta_{jk}\}$,
	then
	$\mathbf{L} = \mathbf{I} - \boldsymbol{\zeta}$, and $ \mathbf{D} = \text{Diag}\{\sigma_1^2, \dots, \sigma_{p}^2 \}$.
	Therefore, $\boldsymbol{\Sigma}^{-1}$ can be further parameterized with the regression coefficients $\boldsymbol{\zeta}$ and residual variances $\{\sigma_{j}^2\}$.

	


	For the fixed effects to be identifiable, the design matrices $\mathbf{U}$ and $(\mathbf{U}_T, \mathbf{W})$ need to be of full rank.
	The covariance matrix $\boldsymbol{\Sigma} $ is identifiable if and only if $\boldsymbol{\Sigma}_{GG}$, $\boldsymbol{\Sigma}_{F_{k}F_{k}} $ and $\boldsymbol{\Sigma}_{GF_k}$ are identifiable for all $k = 1, \dots, K^F$,
	and there exists a $k \in \{1, \dots, K^F\}$,
	such that for some $t \in \{1, \dots, T\}$, the diagonal element $\{\boldsymbol{\Sigma}_{F_{k} F_{k}} \}_{(t,t)}$ is a function of elements in $\boldsymbol{\Sigma}_{F_{k} F_{k}} + \sigma_\epsilon^2 \mathbf{I}_{T \cdot K^F}$.
	For example, if $\{\boldsymbol{\Sigma}_{F_{k} F_{k}}: k = 1, \dots, K^F \}$ are temporal covariance matrices of time series from AR($p$) models with coefficients $(\phi_1^k, \dots, \phi_p^k )$ and $p < T - 1$, then $\boldsymbol{\Sigma}$ is identifiable.

	\section{Parameter Estimation} \label{section: estimation}
	
	Here we delineate the estimation procedure for two groups of parameters of the model in the previous section.
	The first group consists of parameters in characterizing variability in the geometry and functionality of cortical surface images in Section \ref{sec: model for functional surf}, in which the most important are the geometric and functional PC projections (scores),
	and the second group consists of the parameters in the mixed effects model \eqref{eq: mixed model}.

	\subsection{Estimation of the PC Scores}
	
	In practice, the data of each subject are available in a discretized format.
	For each $i$, the geometric shape of cortical surface given by the structural MRI is a triangulated mesh $\mathcal{M}^{\mathcal{T}}_i$ of vertices on the surface as an approximation of the manifold $\mathcal{M}_i$,
	and the function
	$Y_i(t)$ at each time/scan $t$ is represented by a piecewise affine mapping
	$Y^{\mathcal{T}}_i(t): \mathcal{M}^{\mathcal{T}}_i \rightarrow \mathbb{R}$.

	
	In the first step of the estimation, subject-specific geometric shapes $\{\mathcal{M}^{\mathcal{T}}_i \}$ are deformed to a common mean shape (template) $\mathcal{M}^{\mathcal{T}}_0 $
	(using the MATLAB package \textit{fshapesTK} by \cite{Charlier:2015fh}).
	The output is the estimated subject-specific initial momenta $\{\hat{m}^i_{0}\}$ that fully determine the deformations.
	For each subject $i$, $\{\hat{m}^i_{0}\}$ is a group of vectors in $\mathbb{R}^3$ at each vertex of the surface, characterizing the directions and magnitudes to deform the template to the subject-specific surface.
	Then a principal component analysis is applied to the mean-subtracted initial momenta.
	The outputs are the top $K^G$ geometric PCs extracted from the initial momenta and each subject's projection coefficients/scores on the geometric PCs.
	$K^G$ is determined by examining the variance explained by the PCs.
	In addition to subtracting the mean before extracting the PCs, a pre-residualization step can be applied to the estimated initial momenta to regress out the effects of the covariates and confounders.

	In the next step we estimate the functional PCs and projection scores.
	The observed functions $\{Y^{\mathcal{T}}_i(t) \}$ are first mapped to the template domain of $\mathcal{M}^{\mathcal{T}}_0$ with the estimated deformation.
	Specifically, let
	$X^\mathcal{T}_i(t) = Y^\mathcal{T}_i(t) \circ \varphi(\hat{v}_i, \cdot)$,
	where $\{\hat{v}_i \}$ are determined by the estimated initial momenta $\{\hat{m}^i_{0}\}$ obtained in the previous step.
	Then we apply the SM-fPCA algorithm developed by \cite{Lila:2016jj} to the mean-subtracted $\{X^\mathcal{T}_i(t)\}$ to estimate the functional PCs and the associated projection coefficients.
	Pre-residualization can also be applied prior to extracting the functional PCs.
	Additional details in estimating the geometric and functional PCs are available in the Supp. Material.

	\subsection{Estimation of the Mixed Effects Model}
	
	Parameters of interest in the mixed effects model \eqref{eq: mixed model}
	are fixed effects
	$\boldsymbol{\alpha}^G, \boldsymbol{\alpha}^F, \boldsymbol{\beta}$
	and the precision matrix $\boldsymbol{\Sigma}^{-1}$.
	For simplicity of the notation,
	let $p = K^G + T  K^F$.
	Then
	$\mathbf{a} = (\mathbf{a}^G, \mathbf{a}^F) \in \mathbb{R}^{N \times p}$, and
	$\text{vec}(\mathbf{a})
	\sim N(\mathbf{X}  \mathbf{B}, \,
	(\boldsymbol{\Sigma}_{\boldsymbol{\gamma}} + \sigma_\epsilon^2 \otimes \mathbf{I}_p) \otimes \mathbf{I}_{N})$,
	where $\otimes$ denotes the Kronecker product,
	\begin{equation} \label{eq: X}
		\mathbf{X} = \begin{bmatrix}
			\mathbf{I}_{K^G} \otimes \mathbf{U}  &\mathbf{0} \\
			\mathbf{0}  &  \mathbf{I}_{K^F} \otimes (\mathbf{U}_T, \mathbf{W})
		\end{bmatrix}
		\in \mathbb{R}^{(N p) \times (K^G \cdot p_u + K^F \cdot (p_u +p_w))}
	\end{equation}
	and
	$\mathbf{B} =
	\Big( \text{vec}[\boldsymbol{\alpha}^G]^\intercal,
	\text{vec}[(\boldsymbol{\alpha}^F, \boldsymbol{\beta})^\intercal]^\intercal \Big)^\intercal$
	is the re-shaped fixed effects vector.

	Parameters in the mixed effects model \eqref{eq: mixed model} are estimated based on the likelihood function.
	To estimate $\boldsymbol{\Sigma}$, let $ \tilde{\boldsymbol{\gamma}} := (\boldsymbol{\gamma}^G + \boldsymbol{\epsilon}^G, \boldsymbol{\gamma}^F + \boldsymbol{\epsilon}^F)$ denote the random effects combined with the errors, then $\tilde{\boldsymbol{\gamma}}_i \stackrel{\text{iid}}{\sim} N(0, \boldsymbol{\Sigma}) $.
	We will use the parameterization
	$\boldsymbol{\Sigma}^{-1} = \mathbf{L}^\intercal  \mathbf{D}^{-1} \mathbf{L}$,
	$\mathbf{L} = \mathbf{I} - \boldsymbol{\zeta}$
	with $\boldsymbol{\zeta}$ being regression coefficients in
	$\tilde{\gamma}_{j} = \sum_{k=1}^{j-1} \zeta_{jk} \tilde{\gamma}_{k} + \xi_{j}$,
	and
	$\mathbf{D} = \text{Cov}(\boldsymbol{\xi}) = \text{Diag}\{\sigma_1^2, \dots, \sigma_{p}^2 \}$.
	If $\{ \tilde{{\gamma}}_{j} \}$ are observed , elements in $\mathbf{L} $ and $\mathbf{D} $ can be estimated by regressing
	observed $\hat{\tilde{{\gamma}}}_{j}$ on all of $\{\hat{\tilde{{\gamma}}}_{k}: k<j \}$.

	We impose Lasso regularization on the regression coefficients $\{ \zeta_{jk} \}$ for two reasons.
	First, with large $K^G$, $K^F$ and $T$, the number of parameters to estimate in the regressions can easily exceed the number of subjects $N$, making ordinary least squares estimation intractable.
	Second, sparsity of the non-zero regression coefficients is closely related to the structure of the covariance matrix \eqref{eq: Sigma_gamma_tilde}.
	Proposition \ref{prop: sparsity} (with proof in supp. material) states that under mild conditions, the matrix $\mathbf{L}$ preserves the zero blocks in $\boldsymbol{\Sigma}$.
	Essentially, conditions in the proposition describe functional equivalences coming through correlations with the same geometric components.
	Figure 1 in the supp. material exhibits an example of structures of $\boldsymbol{\Sigma}$, $\boldsymbol{\Sigma}^{-1}$ and corresponding $\mathbf{L}$ to demonstrate the results of Proposition \ref{prop: sparsity}.

	\begin{prop} \label{prop: sparsity}
		For $k_1, k_2 \in \{1, \dots, K^F\}$, define the following partial equivalence relation: $k_1 \sim k_2 $ if and only if there exists $g \in 1, \dots, K_G$ such that $\{\boldsymbol{\Sigma}_{G F_{k_1}} \}_{g,\cdot}\neq \mathbf{0}$ and $\{ \boldsymbol{\Sigma}_{G F_{k_2}}\}_{g,\cdot} \neq \mathbf{0}$, where $\{\cdot \}_{g,\cdot}$ denotes the $g$th row of a matrix.
		Suppose $\boldsymbol{\Sigma}^{-1} = \mathbf{L}^\intercal  \mathbf{D}^{-1} \mathbf{L}$ and $\mathbf{L} = \mathbf{I} - \boldsymbol{\zeta}$. Then for $k_1, k_2 \in \{1, \cdots, K^F\}$ and $k_1 < k_2$, $\zeta_{ji} = 0$ for all $i \in \{K^G + (k_1 - 1)T + 1, \dots, K^G + k_1 T \}$ and $j \in \{K^G + (k_2 - 1)T + 1, \dots, K^G + k_2 T \}$ whenever $k_1 \nsim k_2$.
	\end{prop}

	The regression coefficients $\{\zeta_{jk}: k < j\}$ for each $j$ are estimated by minimizing the $L^1$ regularized least squared distance:
	\begin{equation} \label{eq: regression_reg_1}
		{\arg\min}_{\{ (\zeta_{j1}, \cdots, \zeta_{j,j-1}) \}}
		\|\hat{\tilde{{\gamma}}}_j - \sum_{k=1}^{j-1} \zeta_{jk} \hat{\tilde{{\gamma}}}_{k} \|^2
		+ \lambda_j \|\boldsymbol{\zeta}_j \|_1
	\end{equation}
	where $\lambda_j$ is a penalty parameter.
	Values of $\{\lambda_j\}$ can be determined using cross validation. An alternative method of choosing $\{\lambda_j \}$ is to define a vector $\boldsymbol{\tau} = (\tau_1, \dots, \tau_p)$, where $\tau_j$ is the number of non-zero coefficients allowed in the $j$th regression,
	and to select $\lambda_j$ so that
	$\#\{k: \zeta_{jk}\ne 0 \} = \tau_j$.
	The latter is especially suitable for the application of interest, as the number of nonzero elements in the $j$th regressions can be approximated by the desired number of non-zero elements in the $j$th row of the structured covariance matrix \eqref{eq: Sigma_gamma_tilde}, and the latter can be computed given values of $K^G$, $K^F$, $T$, and $j$.
	
	Elements in the diagonal matrix $\mathbf{D}$ are estimated by
	\begin{equation} \label{eq: regression_reg_2}
		\hat{\sigma}^2_j = \frac{1}{N - \#\{k: \zeta_{jk}\ne 0 \} }
		\|\hat{\tilde{{\gamma}}}_j - \sum_{k=1}^{j-1} \hat{\zeta}_{jk} \hat{\tilde{{\gamma}}}_{k} \|^2.
	\end{equation}
	Given the estimates $\hat{\mathbf{D}} = \text{Diag}\{\hat{\sigma}_1^2, \dots, \hat{\sigma}_{p}^2 \}$
	and $\hat{\mathbf{L}} = \mathbf{I} - \hat{\boldsymbol{\zeta}}$,
	we estimate the precision matrix with $\hat{\boldsymbol{\Sigma}}^{-1} = \hat{\mathbf{L}}^\intercal \hat{\mathbf{D}}^{-1} \hat{\mathbf{L}} $
	and the covariance matrix with
	$\hat{\boldsymbol{\Sigma}} = \hat{\mathbf{L}}^{-1} \hat{\mathbf{D}} [\hat{\mathbf{L}}^{-1}]^\intercal $.
	Estimations of coefficients in each regression can be computed separately in a parallel manner,
	making the estimation scalable in high dimensional scenarios.

	\subsection{An Iterative Estimation Algorithm}
	
	We propose an iterative algorithm to estimate the fixed effect coefficients $ \mathbf{B}$ and the covariance matrix $\boldsymbol{\Sigma}$ in the presence of unobserved random effects and errors.
	In the initial step, suppose there are no random effects. The fixed effect coefficients are estimated by regressing $\text{vec}(\mathbf{a})$ on the design matrix $\mathbf{X}$.
	Then the random effects are estimated with residuals of the regression, and the precision matrix is estimated following \eqref{eq: regression_reg_1} and \eqref{eq: regression_reg_2}.
	Given the estimate of $\boldsymbol{\Sigma}^{-1}$,
	the estimated fixed effect coefficients are updated with the generalized least squares estimator (\cite{Lindstrom1990})
	$\hat{\mathbf{B}} = [\mathbf{X}^\intercal (\hat{\boldsymbol{\Sigma}}^{-1} \otimes \mathbf{I}_N) \mathbf{X} ]^{-1} \cdot \mathbf{X}^\intercal (\hat{\boldsymbol{\Sigma}}^{-1} \otimes \mathbf{I}_N) \, \text{vec}(\mathbf{a}) $.
	For the fixed effects, the $L^2$ norm of the difference between estimates in the $n$ and $(n+1)$ steps
	$\Delta(\mathbf{B}^{(n+1)}, \mathbf{B}^{(n)})
	= \|\text{vec}(\mathbf{B}^{(n+1)}) -  \text{vec}(\mathbf{B})^{(n)} \|_2$
	is compared to a given threshold $C_{B} > 0$.
	For the covariance of the random effects, the Kullback-Leibler divergence
	$\Delta_{\text{KL}} (\boldsymbol{\Sigma}^{(n+1)},
	\boldsymbol{\Sigma}^{(n)})
	= \text{tr}\big[(\boldsymbol{\Sigma}^{(n)})^{-1} \boldsymbol{\Sigma}^{(n+1)} \big]
	- \ln |(\boldsymbol{\Sigma}^{(n)})^{-1}
	\boldsymbol{\Sigma}^{(n+1)} |
	- p $
	is compared to a given threshold $C_{\Sigma} > 0$.
	The estimation steps are repeated until the criteria for both of the fixed and random effect parameters fall below the thresholds.
	If further parametrization such as an autoregressive structure is assumed for $\boldsymbol{\Sigma}_{\boldsymbol{\gamma}} = \boldsymbol{\Sigma}_{\boldsymbol{\gamma}}(\boldsymbol{\rho})$, where $\boldsymbol{\rho}$ is a set of unknown parameters, then $\boldsymbol{\rho}$ and $\sigma^2_{\epsilon}$ can be estimated by minimizing the objective
	$\Delta_{\text{KL}} (\boldsymbol{\Sigma}_{\boldsymbol{\gamma}}(\boldsymbol{\rho}) + \sigma^2_{\epsilon}  \mathbf{I}_p, \hat{\boldsymbol{\Sigma}})
	= \text{tr}((\hat{\boldsymbol{\Sigma}}^{-1} (\boldsymbol{\Sigma}_{\boldsymbol{\gamma}}(\boldsymbol{\rho}) + \sigma^2_{\epsilon} + \mathbf{I}_p))
	- \ln |(\hat{\boldsymbol{\Sigma}}^{-1}
	(\boldsymbol{\Sigma}_{\boldsymbol{\gamma}}(\boldsymbol{\rho}) + \sigma^2_{\epsilon} + \mathbf{I}_p) |
	- p $
	over admissible values of $\boldsymbol{\rho}$ and $\sigma^2_{\epsilon}$.

	To test on the fixed effects $\textbf{B}$, we use the Wald-type testing statistic based on the following estimator of the variance of $\hat{\mathbf{B}}$ (see, for example, \cite{Demidenko:2013us}):
	$ [\mathbf{X}^\intercal (\hat{\boldsymbol{\Sigma}}^{-1} \otimes \mathbf{I}_N) \mathbf{X} ]^{-1}$.
	The test statistic for
	$\hat{\mathbf{B}}_i$ is $\hat{\mathbf{B}}_i / \text{diag} ([\mathbf{X}^\intercal (\hat{\boldsymbol{\Sigma}}^{-1} \otimes \mathbf{I}_N) \mathbf{X} ]^{-1} )_i $, which follows the standard normal distribution under the null hypothesis $\mathbf{B}_i = 0$.


	\section{Application to the HCP Cortical Surface Task fMRI} \label{sec: HCP data}
	
	In this section we conduct a comprehensive analysis of the HCP structural MRI and task fMRI with the proposed approach.
	The cortical surface task fMRI was processed with HCP pipeline (as found in ``fMRISurface'' (\cite{Glasser:2013gd}, \cite{HCPmanual}) and
	converted to .csv for statistical analysis using software Connectome Workbench and AFNI package (\cite{Cox:1996jc})).
	For each subject, the processed data consist of a structural MRI scan with the $xyz$ coordinates of $32, 492$ vertices on each hemisphere (left and right), indicating the geometric shape of the cortical surface,
	and a time series of $T=176$ task fMRI scans of BOLD signal values at each of the vertices,
	characterizing the neuronal activation during different stages of the emotion processing tasks.
	In addition, time-invariant age and gender variables and the time series of task stimuli were extracted from the event (EV) files of the data.
	We analyze the effect of time invariant variables and the task stimuli on the geometric shape of the cortical surface and the time series of the task fMRI scans of the full temporal dimension $T=176$.
	Here we focus on analyzing data of the left hemisphere as the majority of the results are symmetric between the left and right hemispheres.
	Previous studies were limited by computational challenges in handling the complete temporal dimension due to the large size of the model components, particularly the covariance matrix. As a result, data were typically analyzed through compression into experimental phases. Our proposed approach overcomes these limitations and allows for the full preservation of temporal information. To highlight the benefits of this approach, we have also conducted an analysis of the data with time collapsed into $T=3$ experimental phases, with the results available in the supplementary material.

	\subsection{Data Analysis}
	The Conte69 cortical surface (\cite{VanEssen:fp}) is used as the template shape in estimating the initial momenta of deformations of subject-specific shapes with the fshapesTK package (\cite{Charlier:2015fh}).
	The principal components (PCs) of the initial momenta are extracted as the geometric PCs (with visualizations in the Supp. material).
	The HCP fMRI data have a correspondence mapping available between vertices on different subjects' surfaces and the template surface, with which we map subjects' functions defined on the subject-specifc shapes to the common template and extract the functional PCs defined on the template.
	Prior to extracting the functional PCs, we averaged each subject's fMRI signals by experiment phases to reduce computation time.
	
	Due to the ultra-high dimensionality of both the geometry and functionality of the cortical surface data, a large number of PCs are required to explain the majority of the geometric (top 30 PCs explains 50\%) and the functional variability (top 20 PCs explains 70\%) in the subjects.
	The variability in the geometric shapes is especially large as all vertices on the hemispheres rather than regions of interest are under examination.
	Here we use a preliminary step to determine the geometric and functional PCs to be included in the mixed effects model.
	In particular, we are interested in PCs that are most relevant to the emotion processing task.
	To this end, as a pre-selection step, we conduct pair-wise Pearson's tests on the correlations between the subjects' projection coefficients on the top 20 geometric PCs and the projections of the fMRI signals during the ``face'' (emotion processing) phase on the top 10 functional PCs.
	Table 1 in the Supp. Material shows the combinations of the functional and geometric PCs sorted in the order of largest to smallest correlations, and the 10 functional PCs and 10 geometric PCs with the largest correlations in the table are included in subsequent analyses.
	
	Age and gender are included in the mixed effects model as time-invariant covariates. 
	There are four categories of age (22-25, 26-30, 31-35, and 36+) and two of gender (female and male). Including the intercept and using the age group of 22-25 and female as  baseline values, the effective number of the time-invariant covariates is $p_u =5$.
	Indicators of task phases during the experiment are included in the model as time-varying covariates. There are three phases: the ``resting'' phase when subjects are in the intermissions between task trials, ``face'' when subjects are asked to match faces with fearful or angry expressions, and ``shape'' when subjects are asked to match pictures of shapes without emotional indications.
	The indicators of the three phases are linear dependent and the resting phase is used as the baseline. Therefore, the effective number of the time-varying covariates is $p_w = 2$.
	
	The mixed effects model for the $i$th subject is
	\begin{align*}
		\mathbf{a}_i^G  &
		= \text{Age}_i \cdot \boldsymbol{\alpha}_{\text{Age}}^G + \text{Gender}_i \cdot \boldsymbol{\alpha}_{\text{Gender}}^G + \boldsymbol{\gamma}_i^G + \boldsymbol{\epsilon}_i^G, \\
		\mathbf{a}_i^F(t)  &
		= \text{Age}_i \cdot \boldsymbol{\alpha}_{\text{Age}}^F + \text{Gender}_i \cdot \boldsymbol{\alpha}_{\text{Gender}}^F + \text{Task}_i(t) \cdot \boldsymbol{\beta} + \boldsymbol{\gamma}_i^F(t) + \boldsymbol{\epsilon}_i^F,
	\end{align*}
	where $\mathbf{a}_i^G$ denote the projection coefficients on the geometric PCs and $\mathbf{a}_i^F(t)$ denote the projection coefficients on the functional PCs at time $t$.
	$\text{Age}_i$ and $\text{Gender}_i$ denote the vector of dummy variables of age and gender categories.
	The effects of the covariates on the geometric PC projections are denoted by $\boldsymbol{\alpha}_{\text{Age}}^G$ and $\boldsymbol{\alpha}_{\text{Gender}}^G$ and the effects on the functional PC projections are denoted by $\boldsymbol{\alpha}_{\text{Age}}^F$ and $\boldsymbol{\alpha}_{\text{Gender}}^F$.
	The time series of task phase indicators are denoted by $\text{Task}_i(t)$ and the their effects on the functional PC projections are denoted by  $\boldsymbol{\beta}$.

	\subsection{Results and Interpretations}
	
	We conduct data analysis in two settings.
	The first is a complete-time setting in which the full time series of $T = 176$ fMRI signals for each subject are preserved.
	Each subject's functional PC projection coefficients at each of the $176$ time points are calculated by projecting the mean-subtracted fMRI signals on the estimated functional PCs.
	The second setting is a compressed setting in which the subjects' time series of fMRI signals are averaged by the $T = 3$ experiment phases prior to estimating the functional PC projection coefficients.

	\subsubsection{Estimation of the Fixed Effects}
	Tables \ref{alpha_g}, \ref{alpha_f} and \ref{beta} show the estimated fixed effects in the $T = 176$ model setting.
	Results of setting $T=3$ are available in the Supp. Material.

	\begin{table}[H]
		\small
		\centering
		\begin{tabular}{lllllllllll}
			\toprule
			& {PC1} & {PC3} & {PC4} & {PC5} & {PC6} & {PC10} & {PC12} & {PC13} & {PC15} & {PC17} \\
			\midrule
			{Gender: M}  & 3.03**       & -1.36.       & -0.56        & 1.64*        & -0.11        & -0.05         & 0.03          & -0.42         & 0.80          & 0.52          \\
			{Age: 26-30} & 0.90         & 0.38         & 0.16         & -0.49        & -0.35        & 0.25          & 0.05          & -0.22         & -0.28         & -0.42         \\
			{Age: 31-35} & 0.67         & -0.48        & -0.34        & 0.19         & -0.21        & 0.02          & -0.07         & -0.84         & 0.04          & -0.12         \\
			{Age: 36+}   & 0.84         & 0.57         & -2.07        & -0.02        & -2.81        & 0.51          & -0.03         & 2.48          & -1.05         & -1.32         \\
			\bottomrule
		\end{tabular}
		\caption{\small Model Setting $T = 176$: Fixed effects $\boldsymbol{\alpha}^G$ of time-invariant covariates (age and gender) on estimated projections associated with the selected geometric PCs. Significance levels: 0.1 (.), 0.05 (*), 0.01 (**), 0.001 ($\dagger$).}
		\label{alpha_g}
	\end{table}

	\begin{table}[H]
		\small
		\centering
		\begin{tabular}{lllllllllll}
			\toprule
			& PC1     & PC2      & PC3      & PC4      & PC5      & PC6      & PC7      & PC8      & PC9      & PC10     \\
			\midrule
			Gender: M  & 1.53$^\dagger$ & -1.66$^\dagger$ & 2.15$^\dagger$  & 0.01     & 0.19$^\dagger$  & -0.36$^\dagger$ & 0.16$^\dagger$  & 0.06*    & -0.43$^\dagger$ & -0.31$^\dagger$ \\
			Age: 26-30 & 0.74$^\dagger$ & -0.90$^\dagger$ & 1.08$^\dagger$  & -0.41$^\dagger$ & 0.32$^\dagger$  & -0.11*   & -0.43$^\dagger$ & 0.55$^\dagger$  & 0.12*    & 0.33$^\dagger$  \\
			Age: 31-35 & 0.57**  & -0.84$^\dagger$ & 1.45$^\dagger$  & 0.15.    & 0.66$^\dagger$  & -0.08    & -0.28$^\dagger$ & 0.69$^\dagger$  & -0.53$^\dagger$ & 0.25$^\dagger$  \\
			Age: 36+   & 0.58    & 4.60$^\dagger$  & -1.73$^\dagger$ & -4.19$^\dagger$ & -3.94$^\dagger$ & 1.70$^\dagger$  & -0.37*   & -1.03$^\dagger$ & -1.76$^\dagger$ & 0.23. \\
			\bottomrule
		\end{tabular}
		\caption{\small Model Setting $T = 176$: Fixed effects $\boldsymbol{\alpha}^F$ of time-invariant covariates (age and gender) on estimated functional PC projections ($\times 10^4$). Significance levels: 0.1 (.), 0.05 (*), 0.01 (**), 0.001 ($^\dagger$).}
		\label{alpha_f}
	\end{table}

	\begin{table}[H]
		\small
		\centering
		\begin{tabular}{lllllllllll}
			\toprule
			& PC1     & PC2      & PC3      & PC4      & PC5      & PC6      & PC7      & PC8      & PC9      & PC10   \\
			\midrule
			{Face}  & 0.51         & 1.71.        & -1.01        & 2.61$^\dagger$      & -0.78        & 1.02*        & -0.97*       & 1.55**       & 1.46$^\dagger$      & 0.60*         \\
			{Shape} & 3.59*        & 1.00         & -0.31        & 1.35*        & 0.55         & 0.66         & 0.31         & 0.66         & 2.02$^\dagger$      & 0.51          \\
			\bottomrule
		\end{tabular}
		\caption{\small Model Setting $T = 176$: Fixed effects $\boldsymbol{\beta}$ of task phases on estimated functional PC projections ($\times 10^2$). Significance levels: 0.1 (.), 0.05 (*), 0.01 (**), 0.001 ($^\dagger$).}
		\label{beta}
	\end{table}

	Comparing the results of the complete time $T = 176$ setting and the collapsed time $T=3$ setting, the estimated fixed effects of the time-invariant covariates on the geometry of the cortical surface are similar in the two settings, which is expected given that the geometry is invariant to task phases and should not be sensitive to different treatments of the tfMRI time series.
	Not many PCs are found to be significantly associated with age and gender, except for PC1 and PC5, which are associated with gender.
	These results are reasonable, as geometric PC1 and PC5 (Figure 2 in Supp. Material) mainly explains the overall size, either inflated or deflated, of the cortical surface comparing to the template, and males are known to have larger brains by volume than females.
	
	On the other hand, for both settings, age and gender have more significant associations with functional PCs than with geometric PCs, indicating more important roles of the demographic factors as well as larger heterogeneity in emotion processing among the population.
	Comparing to the time-collapse setting of $T=3$,
	more significant fixed effects of the time-invariant covariates on the functional PC projections are revealed in the $T = 176$ setting, due to the larger effective sample size and additional information preserved.
	Similar phenomenon can be observed in the estimates of the effects $\boldsymbol{\beta}$ of time-varying covariates, demonstrating improved statistical power of the complete-time model setting. The results here highlight the advantage of the proposed estimation approach given its capability of handling the large dimensionality of the complete time data rather than simply utilizing data of the experimental phases.

	
	\begin{figure}[h]
		\centering
		\begin{subfigure}[b]{0.15\textwidth}
			\includegraphics[width=\textwidth]{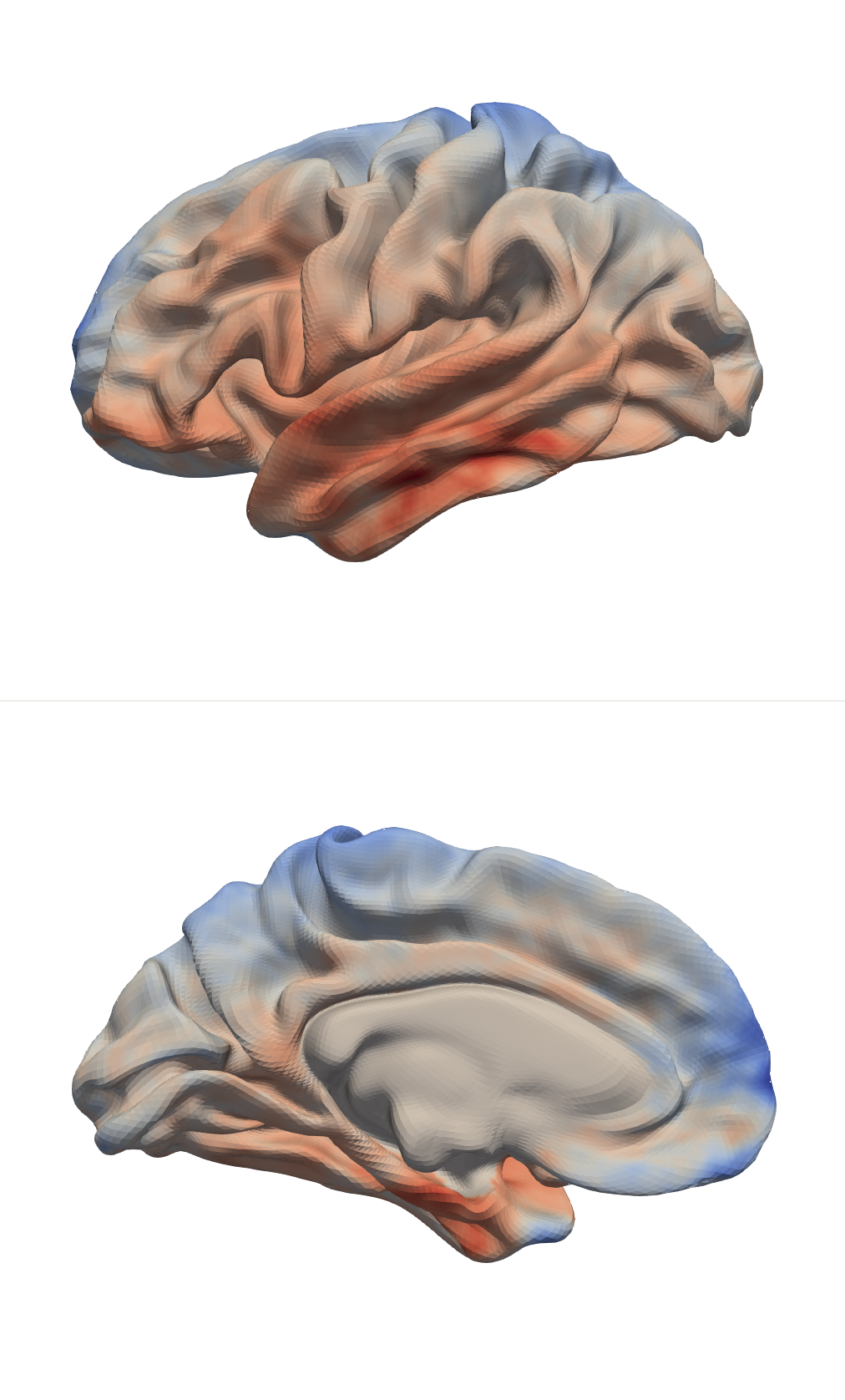}
			\caption{PC4}		
		\end{subfigure}	
		\begin{subfigure}[b]{0.15\textwidth}
			\includegraphics[width=\textwidth]{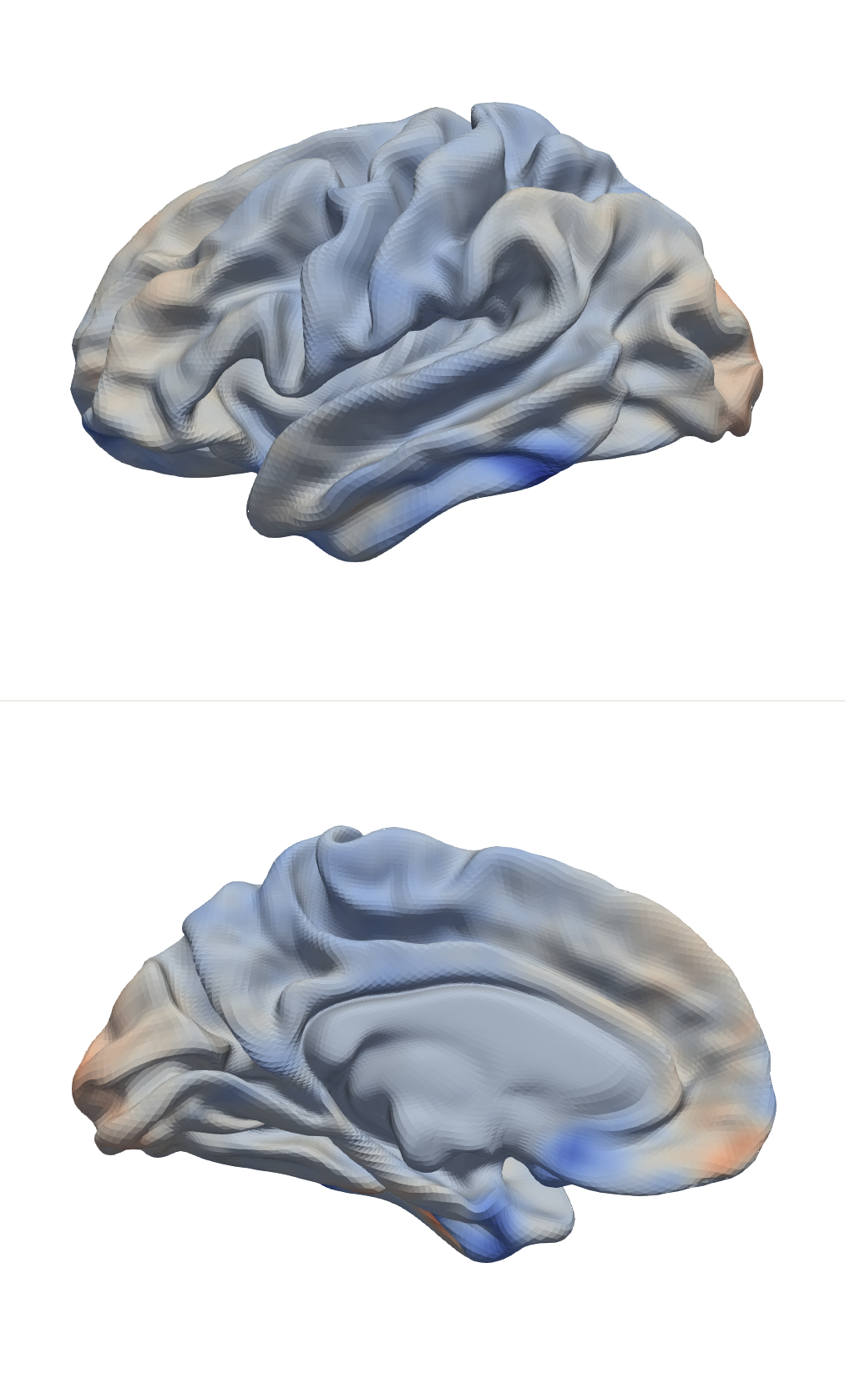}
			\caption{PC6}
		\end{subfigure}
		\begin{subfigure}[b]{0.15\textwidth}
			\includegraphics[width=\textwidth]{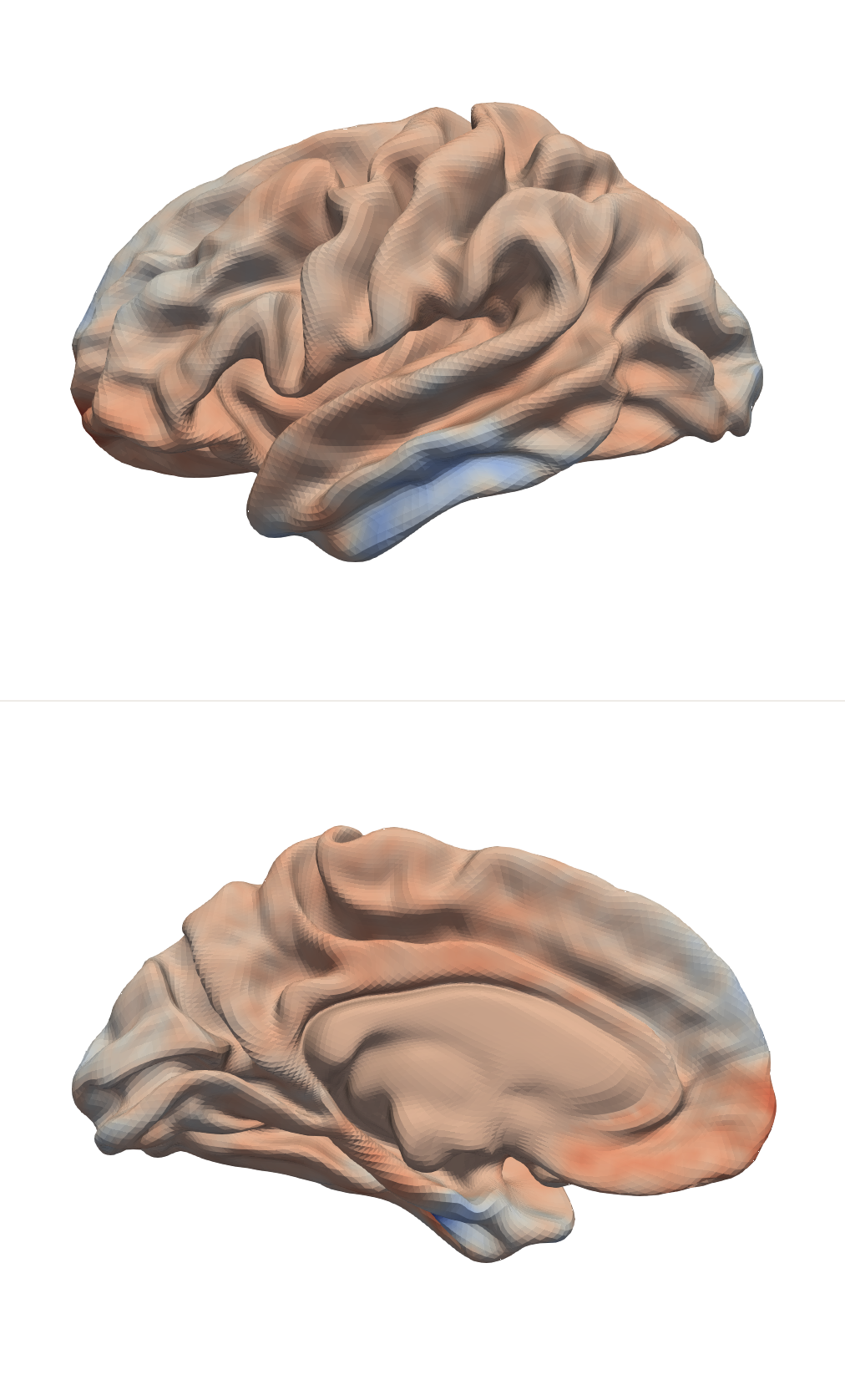}
			\caption{PC7}
		\end{subfigure}	
		\begin{subfigure}[b]{0.15\textwidth}
			\includegraphics[width=\textwidth]{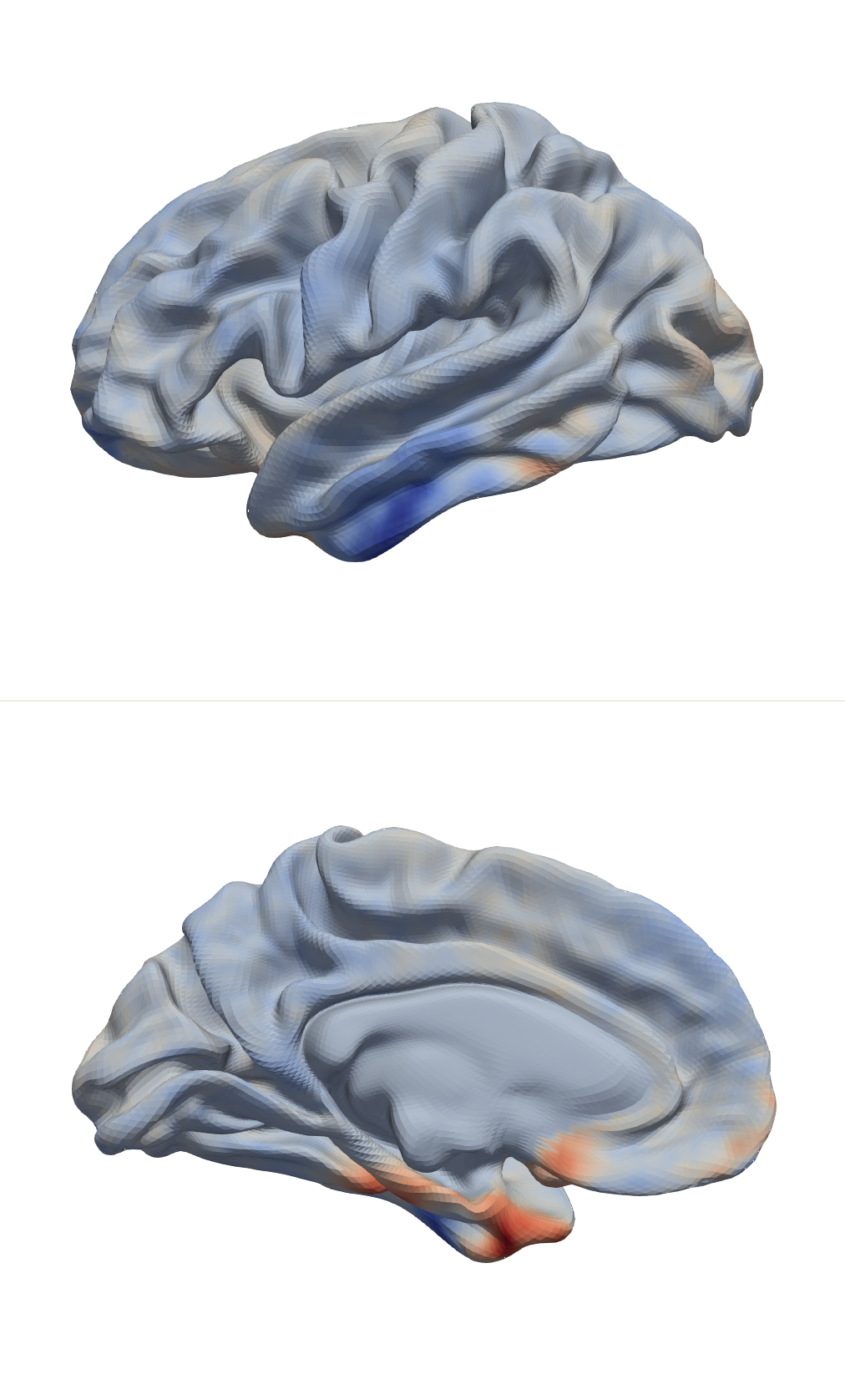}
			\caption{PC8}
		\end{subfigure}	
		\begin{subfigure}[b]{0.15\textwidth}
			\includegraphics[width=\textwidth]{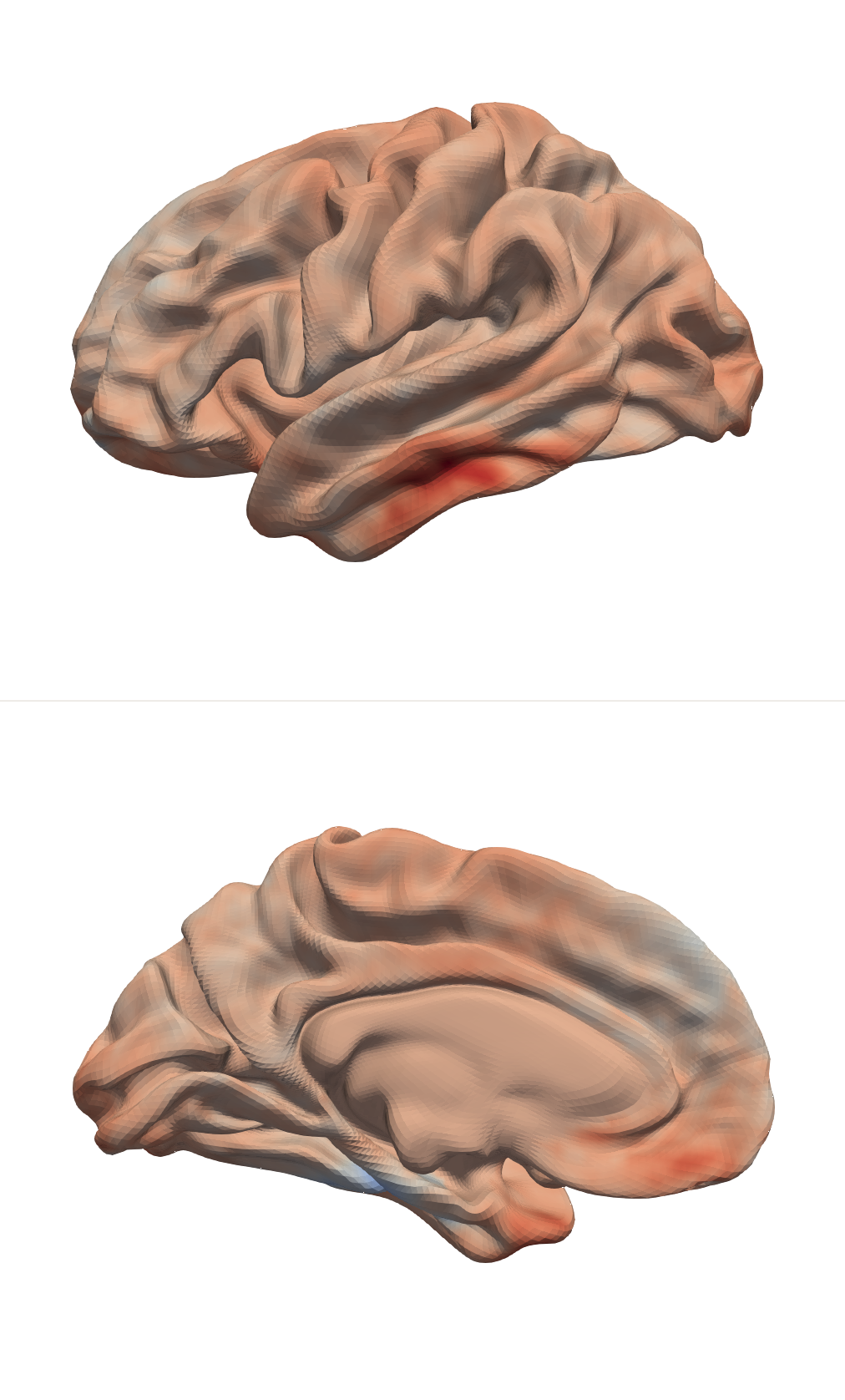}
			\caption{PC9}
		\end{subfigure}
		\begin{subfigure}[b]{0.15\textwidth}
			\includegraphics[width=\textwidth]{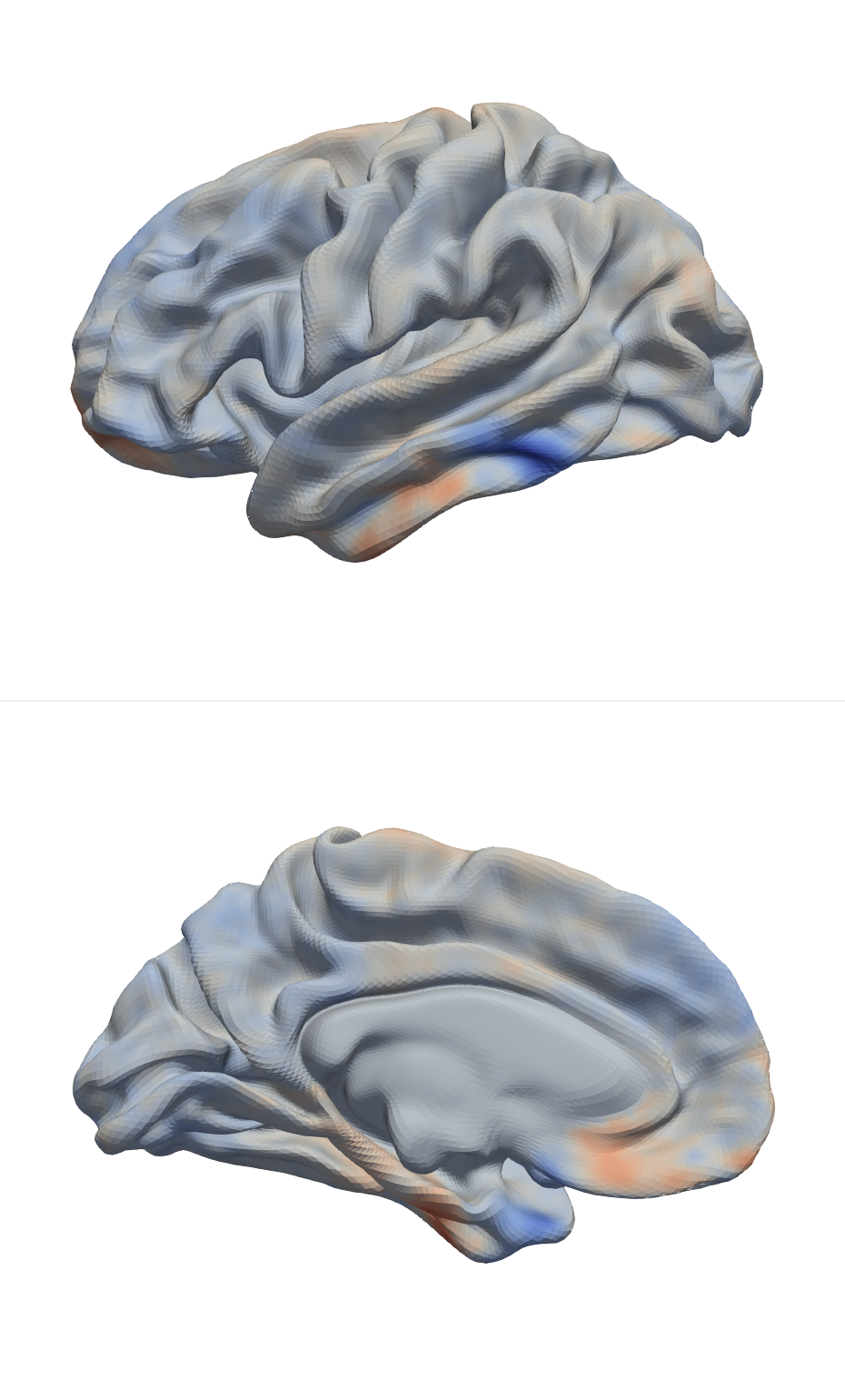}
			\caption{PC10}
		\end{subfigure}
		\caption{Functional PCs associated with the emotion processing task visualized on the template surface. Top: lateral view. Bottom: medial surface. Red color indicates higher values (largest 0.02) and blue color indicates lower values (smallest value -0.02) of the pre-residualized BOLD fMRI signals.}
		\label{figure: functional PC_res}
	\end{figure}

	Figure \ref{figure: functional PC_res} visualizes the functional PCs 4, 6, 7, 8, 9, 10 which are significantly associated with emotion processing task as indicated by the fixed effects of the ``face'' task phase indicator on the corresponding projection coefficients.
	Functional areas associated with the emotion processing task as shown by the deeper colored regions on the PCs are: the inferior and middle temporal, somatosensory and motor cortex, auditory association cortex, orbital and polar frontal cortex, primary visual cortex, parahippocampal, and fusiform.
	Our results supplement the analysis of volumetric subcortical fMRI data by \cite{Barch:2013kqb}, which discovered increased activation of the amygdala extending into the hippocampus, as well as activation in medial and lateral orbital frontal cortices and visual regions including fusiform and ventral temporal cortex during the emotion processing task comparing to the resting state.

	\subsubsection{Estimation of the Random Effects}
	
\begin{figure}[h]
		\centering
		\includegraphics[width=0.95\textwidth]{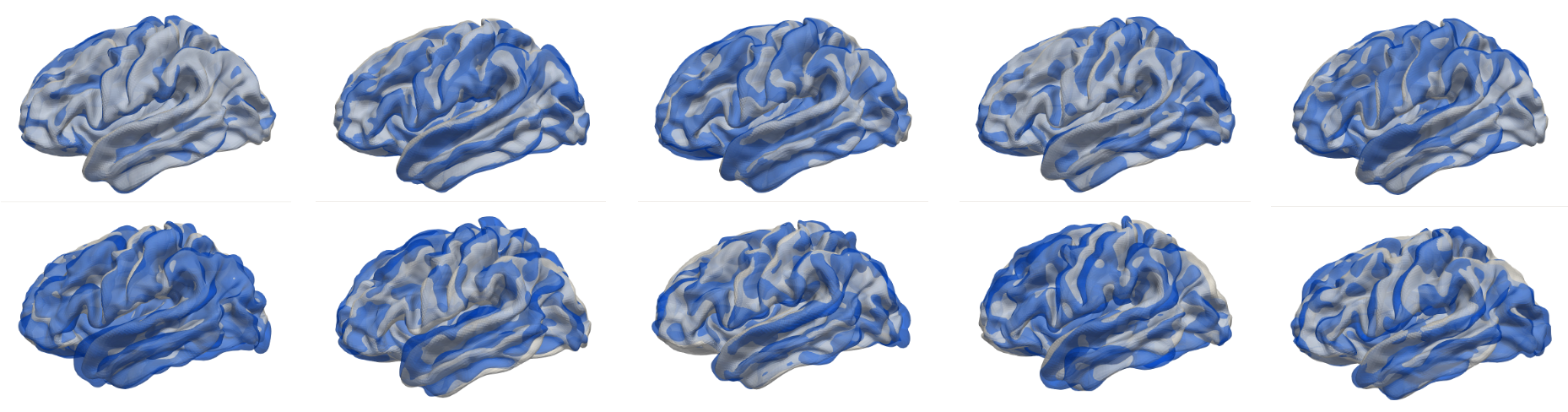}
		\caption{Geometric PCs $6, 10, 12, 15, 17$ associated with functional PCs relevant to emotion processing (lateral view). Top row: deformations of template manifold with geometric PCs' initial momenta $\varphi(c  \psi_j^G, \cdot)$ for $j = 6, 10, 12, 15, 17$ and $c = 1$ (white) and $c = 7$ (blue). Bottom row: deformations of template manifold with $\varphi(c  \psi_j^G, \cdot)$ for $c = -1$ (white) and $c = -15$ (blue).}
		\label{figure: geometric PC_cor}
	\end{figure}

	An important aspect of the results is the covariance matrix of the random effects, which characterizes the relation between the geometry and functionality of the cortical surfaces.
	Figure 3 in the supplementary material shows significant associations between the projections on the geometric and functional PCs via the sparse $\mathbf{L}$ matrix in the Cholesky decomposition $\boldsymbol{\Sigma}^{-1} = \mathbf{L}^\intercal  \mathbf{D}^{-1} \mathbf{L}$.
	A non-zero element in the location corresponding to the $i$th geometric and the $j$th functional PC at time (or experiment phase) $t$ in $\mathbf{L}$ suggests a potentially importance correspondence between the PCs.
	Figures \ref{figure: geometric PC_cor} visualizes geometric PCs $6, 10, 12, 15, 17$, which were found to be associated with the functional PCs 4, 6, 7, 8, 9, 10 (Figure \ref{figure: functional PC_res}) during the emotion processing task phase.
	We plot the manifolds that represent shapes of the cortical surfaces associated with these geometric PCs
	by applying deformations $\varphi(c \cdot \psi_j^G, \cdot)$ to the Conte69 template, where $\psi_j^G$ denotes the vector field associated with initial momenta of the $j$th geometric PC.  
	To visualize how the shapes corresponding to different geometric PCs vary from the template, for each geometric PC we plot the cortical surfaces deformed from the template following the initial momenta of the PC, multiplied by different scalers $c = -15, -1, 1, 7$.
	
	
	\begin{figure}[h]
		\centering
		\includegraphics[width=0.4\linewidth]{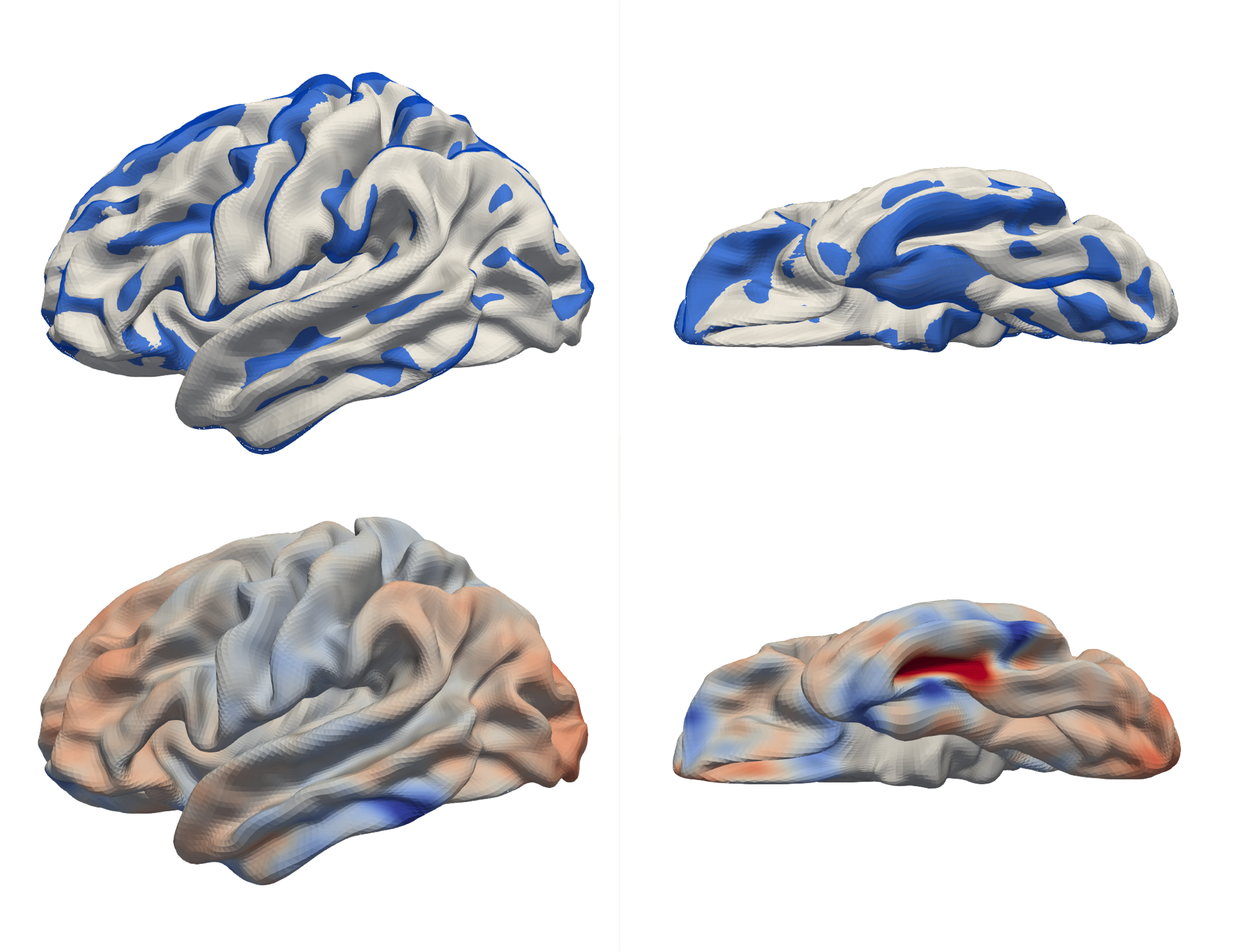}
		\caption{A comparison of geometric PC6 (top) and functional PC6 (bottom)}\label{fig: F6G6}
	\end{figure}
	
	A particular interesting finding is in
	Figure \ref{fig: F6G6}, which provides a side-by-side comparison of functional PC 6 and geometric PC 6.
	Regions of activation marked by deep red and blue colors on the functional PC (bottom), especially near the inferior temporal and fusiform gyri, have a visible correspondence with the geometric shape (top) of these regions:
	the blue/white colors on the top right panel, which indicate inflated/deflated gyri, are matched with the deep blue/red colors on the bottom right of activated areas.
	The occipitotemporal sulcus in white (top right panel) also has a visible correspondence with the deep red color in the functional PC colormap (bottom right panel).
	A similar correspondence can be seen in the temporal lobe area.
	These results are interesting as they hint potential associations between the anatomical development and functionalities of regions on the cortical surface, specifically with respect to the processing of emotions. They also indicate more generally that there are links between the shape of regions and how they functionally behave.

	\section{Concluding Remarks}
	
	In this paper we propose a framework for jointly modeling the geometric and functional variability in functional surfaces including the structural MRI and task fMRI of the cortical surface.
	We characterize effects of subject-specific covariates and exogenous stimuli on both the geometry and functionality while accounting for their mutual-influences with a unified mixed effects model.
	We develop a computationally efficient estimation method for the proposed mixed effects model by iteratively estimating the fixed effects and elements in the Cholesky decomposition of the precision matrix of random effects.
	In particular, elements in the rows of the Cholesky decomposition are estimated via regularized regressions to circumvent the computational burdens in dealing with high dimensional covariance matrices.
	The proposed method is scalable and automatically guarantees the positive-definiteness of the estimated precision and covariance matrix.
	
	We apply the proposed method to the HCP cortical surface structural MRI and task fMRI data in the complete time setting with $T=176$ fMRI scans and compare the results to those from the $T=3$ aggregated experiment phases.
	We analyze the fixed effects of age and gender and task stimuli on both geometry and functionality of subjects' cortical surface as well as the covariance between the geometry and functionality.
	The proposed approach reveals patterns of geometric shapes and activated regions associated with the covariates, as well as unique modes of correspondence between the shapes of the cortical surface and functionality related to emotion processing.
	In the appendix, we further examine the performance and computational efficiency of the proposed method with a comprehensive simulation study using synthetic data and demonstrate its advantage when compared to other methods including REML.
	While we focus on the analysis of vertex-wise cortical surface data, the model and method proposed here can be readily applied to study the geometry and functionality of regions of interest (ROI) of the cortical surface.
	The proposed model and method can also be applied beyond cerebral cortical surfaces and used for functional surfaces in general, such as the study of bone surface in \cite{Gee:2018hj}.
	
	There are several possible future directions for us to pursue.
	First, as in many studies on functional data, the alignment of subject-specific functions to a common template domain induces identifiability issues, which will be carried over to statistical analyses of the aligned functions. 
	The reason is there are often many ways to align/register the subject-specific domains and each way can result in a different function on the common template and thus affect the inference results.
	This identifiability issue is alleviated in our study, as the proposed joint model is able to capture the variability in both geometry and functionality of the functional surfaces.
	A possible approach in future studies for improving the alignment of the functions is to incorporate prior knowledge on functional regions, such as functional atlas or ROIs as well as landmarks, into the registration step.
	
	Second, in modeling the variability in functionality, an alternative approach is to use the multilevel PCA (\cite{Di:2009dz}) to further separate the inter- and intra-subject (temporal) variability of the aligned tfMRI signals. The PC projections at each level can be used to study the effects of covariates on neuronal activation at different phases of the tasks.
	
	Third, in the proposed model and estimation method we do not impose extra constraints on the elements of the covariance/precision matrix other than the regularization on elements in the Cholesky decomposition of the precision matrix. As a result the estimation of the covariance matrix is flexible and adapts well to a wide range of covariance structures.
	If prior knowledge implies further constraints on the covariance matrix, elements in the Cholesky decomposition can be re-parameterized to satisfy, at least approximately, the desired structure of the covariance matrix.
	
	Finally, if multiple modalities of the functional surfaces are available for the subjects, such as different types of fMRI including resting state and task fMRI, and Diffusion Tensor Imaging (DTI), the proposed model can be extended to characterize the correlations among different modalities while examining the covariates' effect on individual modalities.

	\appendix
	
	\section*{Appendix: Simulation Studies} \label{sec: simulations}
	
	We further demonstrate the effectiveness of our proposed approach through simulations. In this study, we analyze synthetic functional manifolds that are similar to the structural and task fMRI data obtained from the cortical surface, but with a reduced dimensionality to minimize computation and allow for testing of various methods, including existing methods not suited for high-dimensional cortical surface data, under different conditions.

	\section{Generation of the Functional Manifolds}

\begin{figure}[h]
		\centering
		\includegraphics[width=0.8\linewidth]{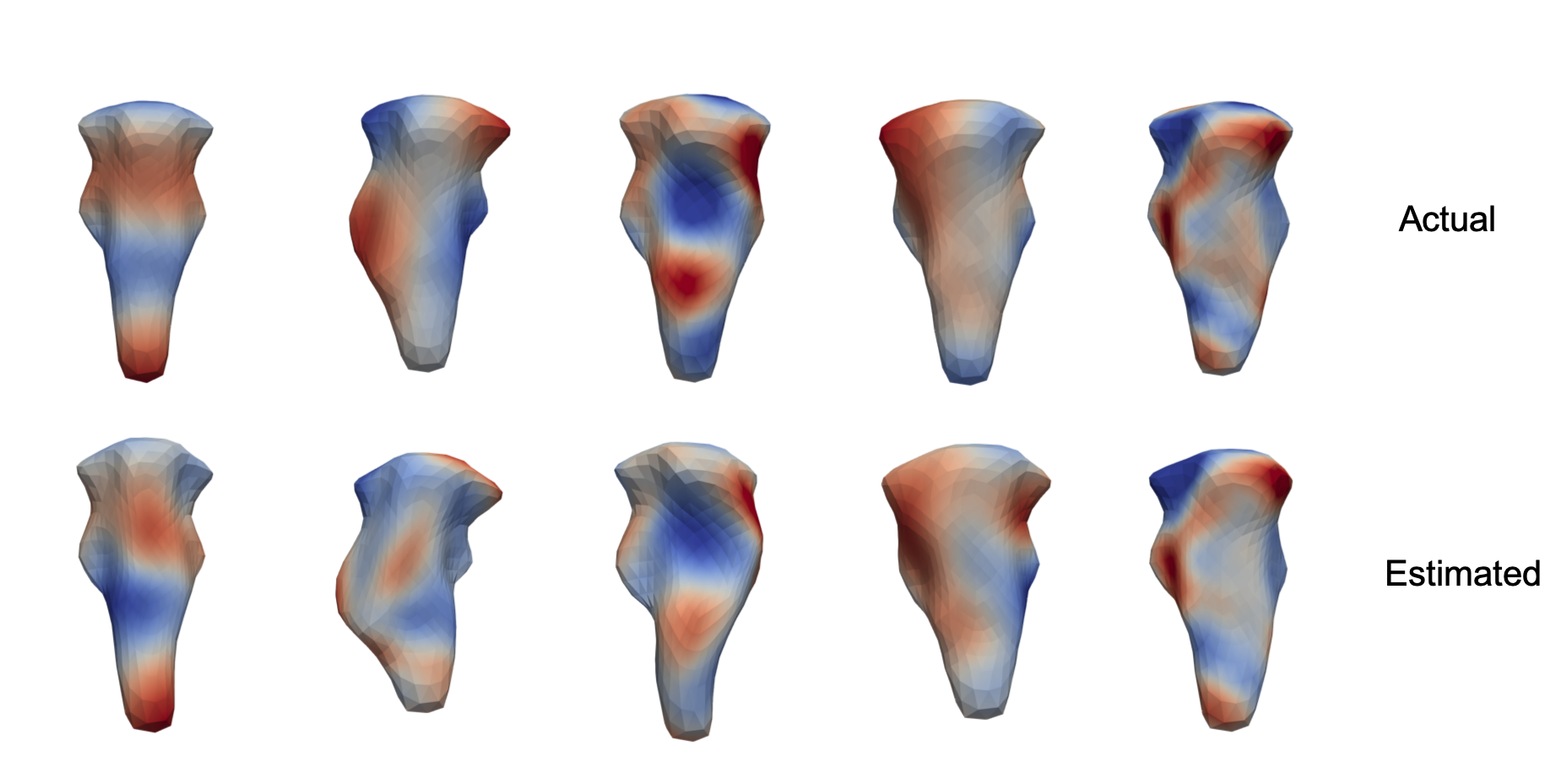}
		\caption{The geometric and functional PCs visualized by overlaying colormaps of functional PCs on shapes of geometric PCs with the same indices. Top row: the PCs in the generative model. Bottom row: estimated PCs from one simulation run of 200 subjects.}
		\label{fig: PCs}
	\end{figure}

	We first create $K^G$ geometric PCs and $K^F$ functional PCs
	using a template manifold $\mathcal{M}_0$ resembling the human brain stem created by \cite{Lila:2016jj} (visualization of the template shape available in Supp. Material).
	To generate shapes associated with geometric PCs, we create $K^G$ orthogonal deformation momenta to map $\mathcal{M}_0$ to manifolds associated with the PCs.
	The functional PCs are mappings from $\mathcal{M}_0$ to $\mathbb{R}$ and are generated to have correspondence with the geometric PCs.
	For the $k$th functional PC, large values (red) are generated on vertices where the shape of the $k$th geometric PC is inflated, and smaller values (blue) are generated on vertices where the $k$th geometric PC are deflated.
	The generated PCs are visualized on the top row of Figure \ref{fig: PCs}, in which the shapes represent the geometric PCs while the colormap on top of the shapes represents the functional PCs.

	The simulation study is conducted under a small temporal-dimension setting with $T = 5$ and a large temporal-dimension setting with $T = 50$.
	For each setting we run $20$ independent simulations.
	In each simulation run, $N = 200$ subjects' functional surfaces are generated from the $K^G = 5$ geometric and $K^F = 5$ functional PCs in Figure \ref{fig: PCs}.
	Each subject's projection coefficients $\mathbf{a} = (\mathbf{a}^G, \mathbf{a}^F)$ on the geometric and functional PCs are generated as the outcomes of the mixed effects model \eqref{eq: mixed model} in the main text.
	For the mixed effects model $p_u = 2$ time-invariant covariates are sampled independently from $U_1 \sim N(3, 3)$ and $U_2 \sim t_3 + 3$.
	In addition, $p_w = 2$ time-varying signals $\mathbf{W}$ are generated for all subjects to resemble the stimuli in the task fMRI (see the Supp. Material for plots of the signals).
	For the fixed effects coefficients,
	we select $\text{vec}(\boldsymbol{\alpha}^G) = (1, 0, 0, 1, 0.5, 0.5, 0.2, 0.6, 1.5, 0.5)$ for effect sizes of $\mathbf{U}$ on the geometric PC projections,
	$\text{vec}(\boldsymbol{\alpha}^F) = (1, 1, 0.5, 1.5, 0.5, 0.5, 1, 0, 0, 1)$ for effect sizes of $\mathbf{U}$ on the functional PC projections,
	and $\text{vec}(\boldsymbol{\beta}) = (0, 1, 1, 0, 0.5, 0.5, 1.5, 0.5, -0.3, -0.7)$ for effect sizes of $\mathbf{W}$ on the functional projections.
	
	The random effects $\boldsymbol{\gamma}$ in the mixed effects model are sampled independently from
	$N(\mathbf{0}, \boldsymbol{\Sigma}_{\boldsymbol{\gamma}} )$,
	where $ \boldsymbol{\Sigma}_{\boldsymbol{\gamma}} $ is constructed using the Cholesky decomposition \eqref{eq: decomp_precision}.
	The constructed $\boldsymbol{\Sigma}_{\boldsymbol{\gamma}}$ has blocks $\boldsymbol{\Sigma}_{GG}  = \text{Diag}(25,16,9,4,1)$ and each of $ \{\boldsymbol{\Sigma}_{F_k F_k}: k = 1, \dots, 5\}$ following approximately the covariance matrix of an AR(1) model with variances $(30, 20, 10, 5, 1)$, respectively.
	The blocks $\boldsymbol{\Sigma}_{G F_k}$ are constructed to contain significantly non-zero values to represent the covariance between the geometry and functionality.
	Finally, errors $\boldsymbol{\epsilon} \stackrel{iid}{\sim} N(0, \sigma_{\epsilon}^2)$ with $\sigma_{\epsilon} = 0.5$ are generated independently of the random effects.
	In constructing the covariance matrices, we do not impose further parametrization on $ \boldsymbol{\Sigma}_{\boldsymbol{\gamma}} $, and thus the covariance matrix is identifiable up to $ \boldsymbol{\Sigma} = \boldsymbol{\Sigma}_{\boldsymbol{\gamma}} + \sigma_\epsilon^2 \mathbf{I}$.
	
	The deformation operator $\phi_i$ for mapping the $i$th subject's shape from the template $\mathcal{M}_0$ is calculated as the linear combination of the geometric PCs with the projection coefficients $\mathbf{a}_i^G$ as the weights.
	Similarly, the function  of the $i$th subject at each time $t$, defined on the domain of $\mathcal{M}_0$, are obtained by calculating the linear combinations of the functional PCs weighted by $\mathbf{a}_i^F(t)$ .
	Finally, the observed function of the $i$th subject, defined on the subject-specific domain $\mathcal{M}_i$, is given by the composite of $\phi_i$ and the function on $\mathcal{M}_0$.

	The propose estimation approach is applied to the generated synthetic functional manifolds.
	To examine the effectiveness of the proposed estimation for the mixed effects model,
	we also estimate the mixed effects model with actual values of $ (\mathbf{a}^G, \mathbf{a}^F) $ as the outcome to separate errors in the mixed effects model estimation from those induced in estimating the deformations and extracting the PCs.

	We compare results using the proposed method with two alternative methods for mixed effects model estimation, whenever the alternatives are computationally feasible.
	The first is to use an iterative algorithm similar to the proposed but without any regularization on the elements in the Cholesky decomposition of $\boldsymbol{\Sigma}^{-1}$.
	The second is to use restricted maximal likelihood estimation (REML). Here we implement a modified REML algorithm based on the R package \textit{lme4} (\cite{Bates:2015cf})
	In the modified REML estimation, we assume the structure of $\boldsymbol{\Sigma}$ including the zero blocks is known \textit{a priori} to avoid the computational burden in optimizing over the complete set of $p(p+1)/2$ parameters.
	Not all parameter settings guarantee the positive definiteness of the covariance matrix, without which the REML cannot proceed.
	To address this issue we implement truncations in which non-positive eigenvalues of the covariance matrix given the proposed parameter setting are replaced with $ 10^{-5}$ so that the resulting matrix is positive definite.
	We also estimate the fixed effects without accounting for the random effects and include the results in the comparison.

	\section{Estimation Results}
	
	\subsubsection{$K^G = 5, K^F = 5, T = 5, N =200$} \label{sec: T5}
	
\begin{figure}[H]
		\centering
		\begin{minipage}[b]{0.48\textwidth}
			\centering
			\includegraphics[width=\textwidth]{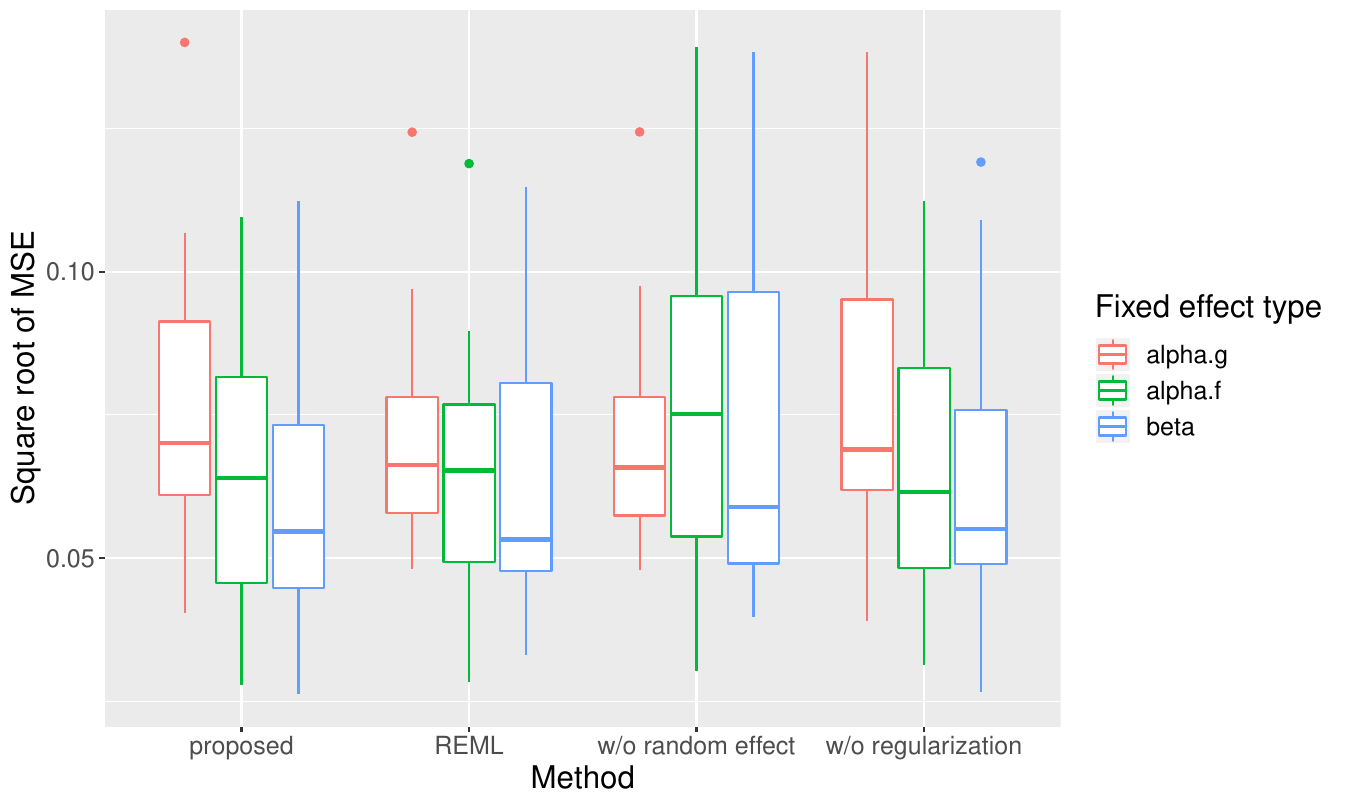}
		\end{minipage}
		\hfill
		\begin{minipage}[b]{0.48\textwidth}
			\centering
			\includegraphics[width=\textwidth]{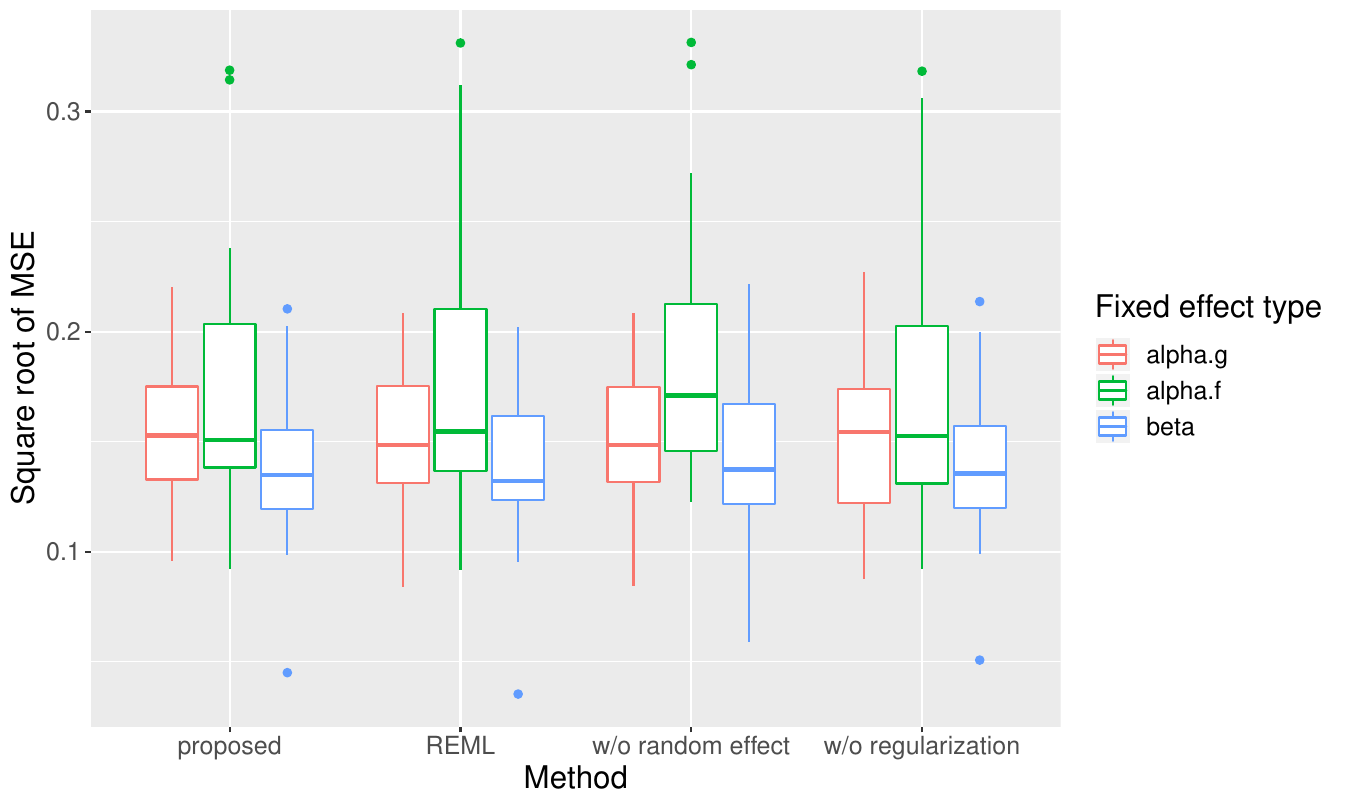}
		\end{minipage}
		\caption{Estimation of fixed effects, grouped by methods and fixed effects type. Left: with actual PC projection coefficients. Right: with estimated PC projection coefficients.} \label{fig: fixed_T5}
	\end{figure}

	Figure \ref{fig: fixed_T5} summarizes the estimation of the fixed effects via box plots. The left panel shows the square roots of the mean squared errors (MSE) from all simulation runs in which the estimates are obtained with the actual PC projection coefficients.
	Results are averaged by fixed effect types: $\boldsymbol{\alpha}^G$, $\boldsymbol{\alpha}^F$ and $\boldsymbol{\beta}$,
	and methods (grouped boxplots from left to right): proposed, REML, estimation without accounting for random effects, and estimation without imposing regularization on the covariance matrix.
	The right panel shows the results obtained from the estimated PC projection coefficients.

	With the actual values of PC projections $\mathbf{a}$, different methods have similar performance in estimating $\boldsymbol{\alpha}^G$, except for the procedure without imposing regularization on the covariance matrix, which leads to larger errors.
	In estimating $\boldsymbol{\alpha}^F$ and $\boldsymbol{\beta}$, the procedure without accounting for random effects under-performs comparing to the other methods.
	The proposed method and REML have the best performance, which is not surprising, as the other two methods either over-simplifies the covariance matrix as the identity matrix or fails to recognize the sparsity in the covariance structure.
	The right panel of Figure \ref{fig: fixed_T5} demonstrates to what extent the fixed effects estimation is affected by the biased introduced in the geometric registration of the manifolds and the estimation of the PCs. The median errors nearly doubled compared to the estimates with the actual $\mathbf{a}$, and the comparison of performances of different methods is similar to that with the actual $\mathbf{a}$.

	
	\begin{figure}[h]
		\centering
		\includegraphics[width=1\linewidth]{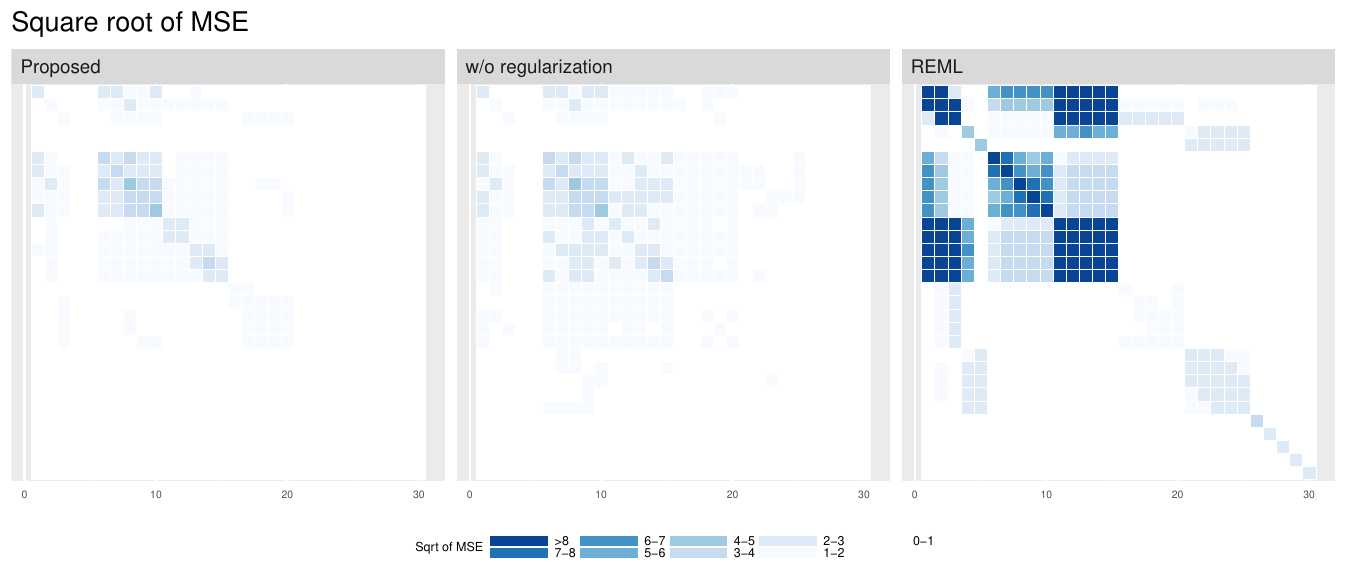}
		\caption{Square root of MSE of estimated covariance matrices using different methods (estimates obtained with actual PCs). }
		\label{fig: sigma_MSE_T5_0}
	\end{figure}
	
	Figure \ref{fig: sigma_MSE_T5_0} compares estimation of the covariance matrix with the actual PC projection coefficients with the square roots of the MSE in each element.
	The proposed method is able to recover the structure of the covariance matrix with high precision, including the temporal covariance of the time series $\mathbf{a}^F$ and the covariance between the geometric and functional PC projections (color bands on the top left corner of the matrix).
	The procedure without regularizing the covariance matrix is able to recover most of the covariance structure, but has many false recoveries of non-zero elements in the matrix.
	REML performs much worse than the aforementioned two methods, suffering from large errors in estimating the non-zero elements in the covariance matrix.
	The MSE of REML in Figure \ref{fig: sigma_MSE_T5_0} contains apparent color blocks, indicating REML fails to recover most of the covariance structure, despite the fact that the positions of zero elements in the matrix are already given ``for free'' in the simulations.
	It is also worth noting that REML's performance relies heavily on the initial value.
	Additional figures are available in the Supp. Material for estimation results of the covariance matrix obtained with the estimated PC projection coefficients.
	Although less accurate due to errors induces in the estimation of deformation and PCs,
	the proposed method still has satisfactory performance comparing to the other methods, which suffer from false recoveries of non-zero elements in the covariance matrix.

	Regarding computational efficiency,
	the proposed method took about 8 seconds to finish the estimation, the method without regularization took 6.8 seconds,
	and the REML took 13.5 minutes.
	Here the REML method is given an advantage of knowing the zero blocks in the covariance matrix, a much reduced number of parameters needs to be optimized over.
	In more practical scenarios, the estimation with REML is expected to take a longer time.
	

	\subsubsection{$K^G = 5, K^F = 5, T = 50, N = 200$}
	
	We also examine a large-$T$ scenario that reflects the realistic dimension of cortical surface fMRI data.
	Same parameters are used in generating the functional manifolds.
	
	We also apply the ``no regularization on the covariance matrix'' method to compare with the proposed method.
	With $T=50$,
	elements in the covariance matrix can no longer be estimated with the regression method due to over-fitting when the number of elements exceeds the number of subjects. Here we modify the estimation algorithm: in each iteration the covariance matrix is estimated with a truncated empirical covariance matrix whose non-positive eigenvalues are replaced with a small positive constant.
	In this large $T$ setting, the REML becomes computationally infeasible and thus no results are available for the comparison.

\begin{figure}[H]
		\centering
		\begin{minipage}[b]{0.48\textwidth}
			\centering
			\includegraphics[width=\textwidth]{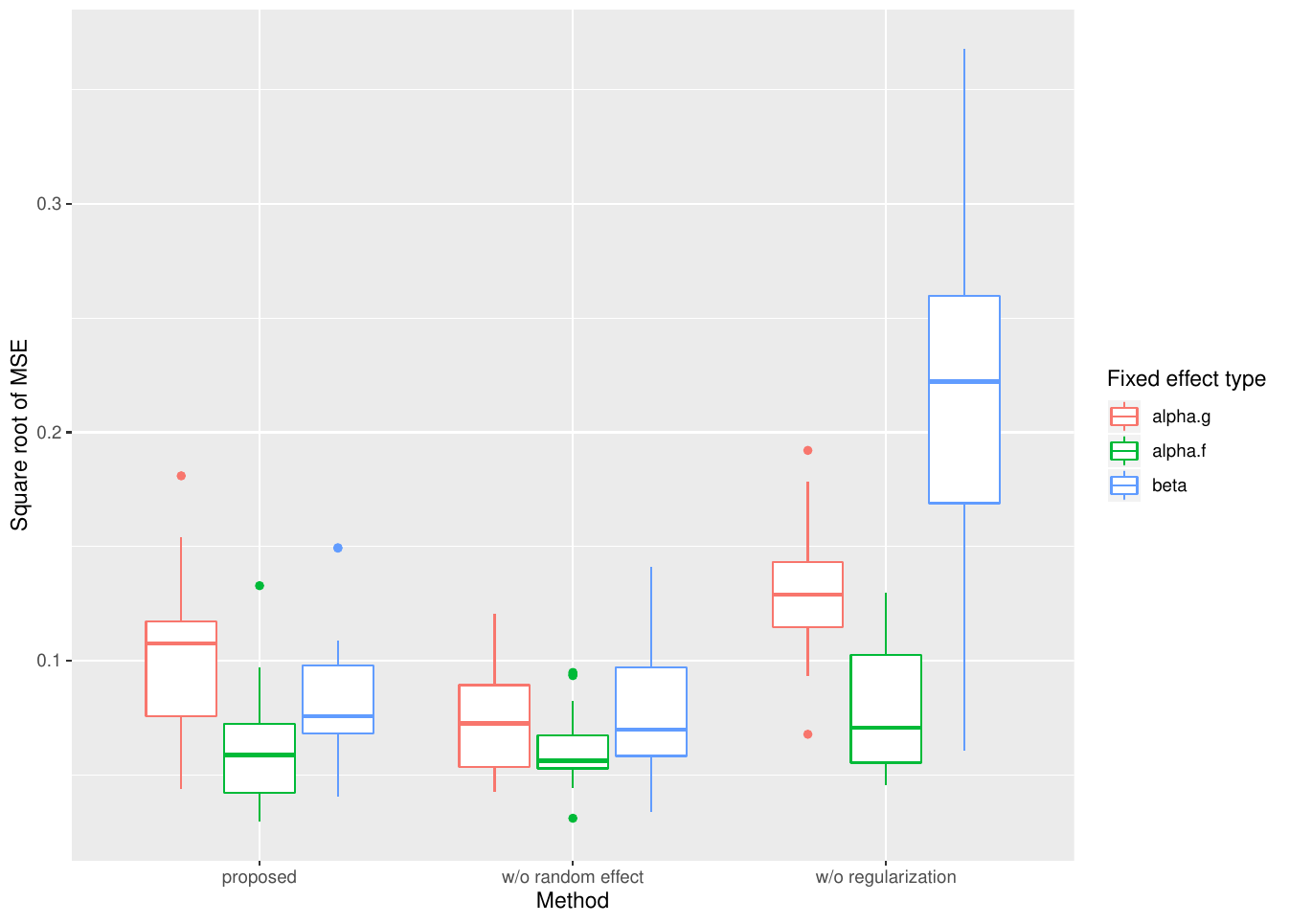}
		\end{minipage}
		\hfill
		\begin{minipage}[b]{0.48\textwidth}
			\centering
			\includegraphics[width=\textwidth]{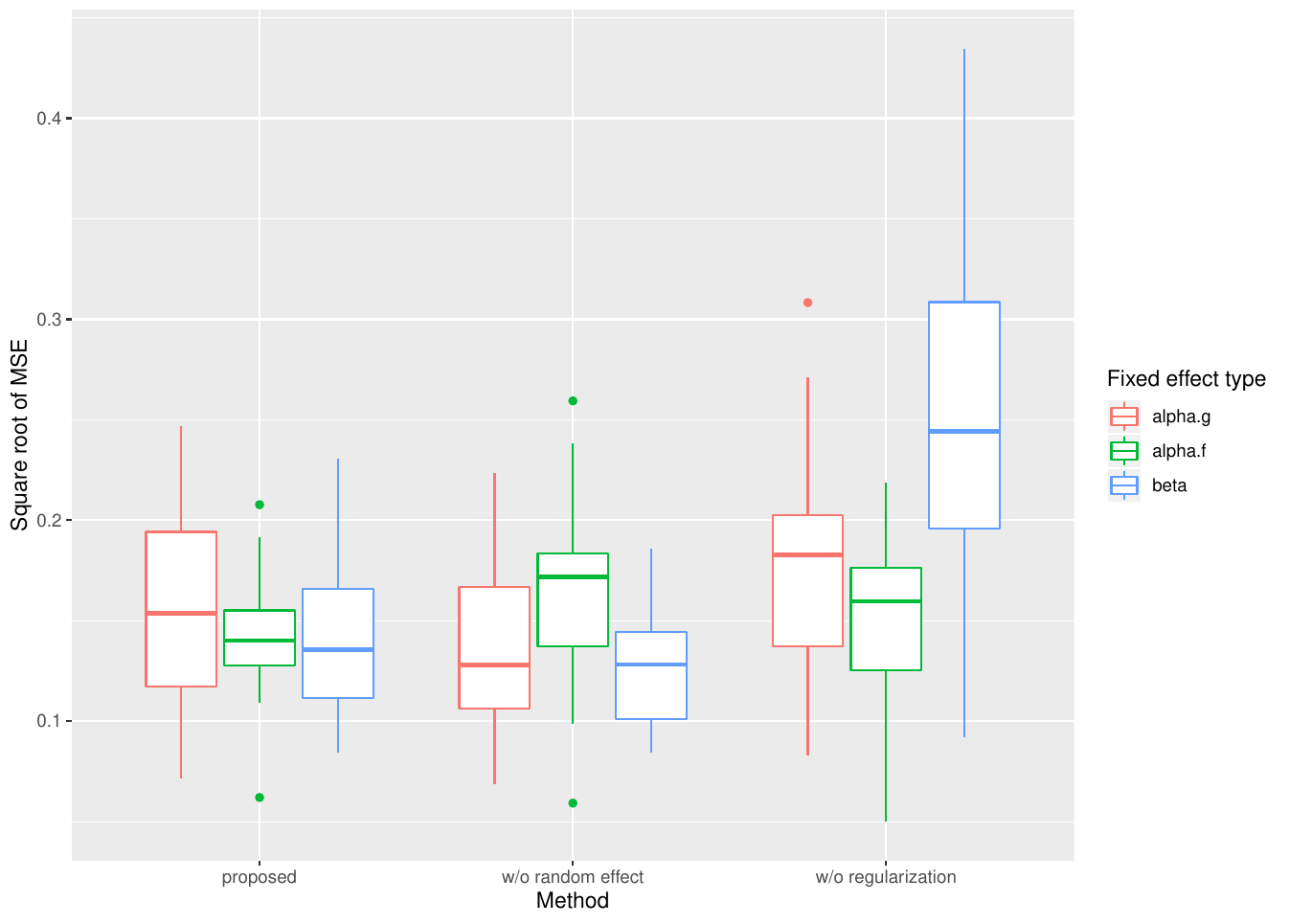}
		\end{minipage}
		\caption{Estimation of fixed effects, grouped by methods and fixed effects type. Left: with actual PC projection coefficients. Right: with estimated PC projection coefficients.} \label{fig: fixed_T50}
	\end{figure}

	The left panel of Figure \ref{fig: fixed_T50} summarizes estimates of fixed effect coefficients via boxplots grouped by fixed effect types and methods.
	It is apparent that the fixed effects estimation is heavily affected when the covariance matrix is estimated poorly with the ``without regularization'' method, which is expected as the empirical covariance matrix performs poorly in the high dimensional scenario.
	Estimation with the proposed method is comparable to the method without accounting for random effects.
	The right panel of Figure \ref{fig: fixed_T50} displays the estimation results with estimated PC projections. Due to the additional errors induced during the geometric registration and the PC estimation, the errors increases by about 50\% for all three methods.

	\begin{figure}[H]
		\centering
		\includegraphics[width=0.75\linewidth]{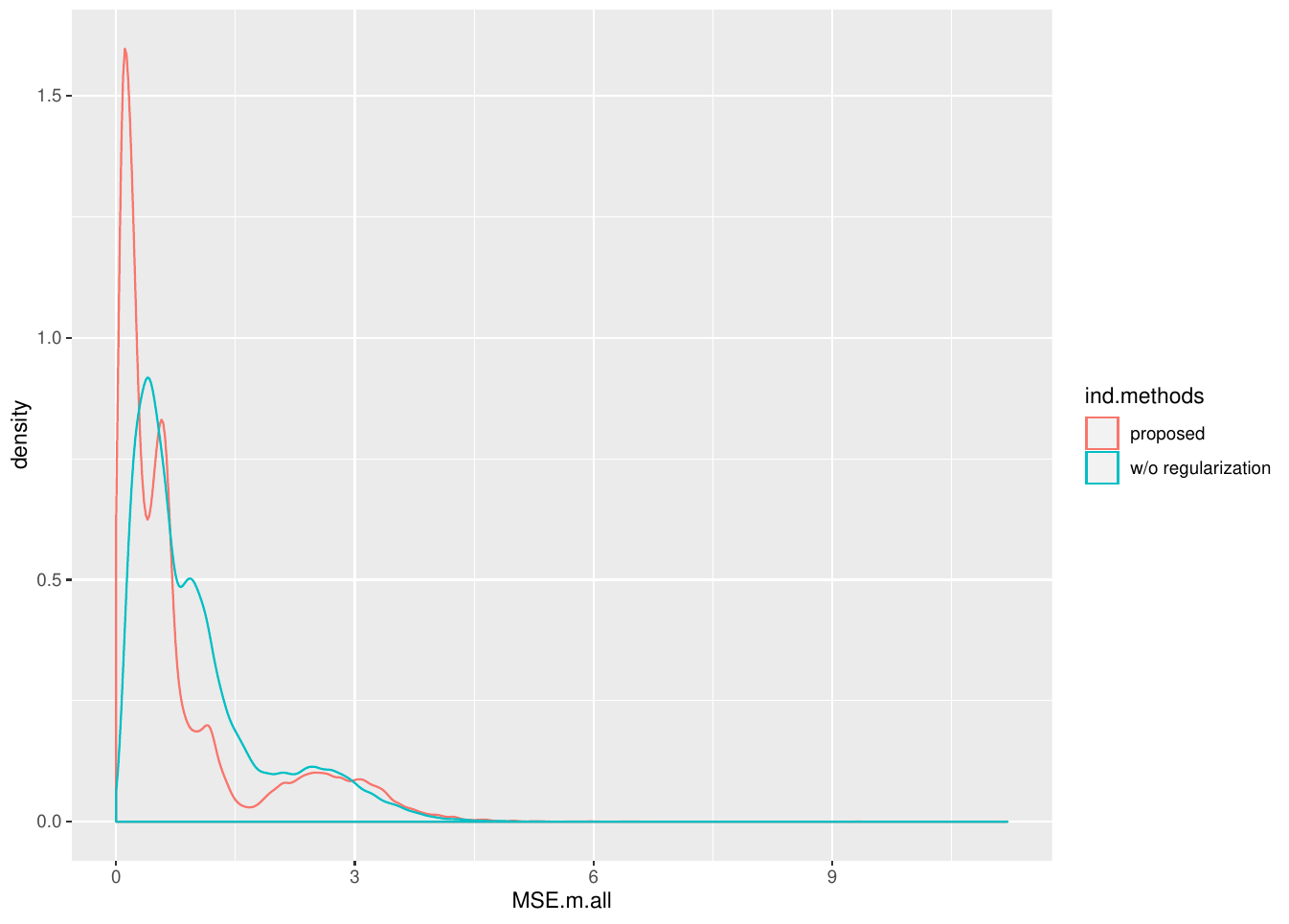}
		\caption{Density curves of square roots of MSE for each element in the covariance matrix.}
		\label{fig: MSE_hist_T50_1}
	\end{figure}

	Due to the large dimensionality of the covariance matrix, instead of showing the estimated covariance matrix averaged over the simulations, we show in Figure \ref{fig: MSE_hist_T50_1} the density curves of square roots of MSE for each element in the covariance matrix. The density curve of proposed method peaks at near zero and has the majority of mass concentrated in $[0, 1.5]$. The curve of the ``no regularization'' method shifts towards large values of the error, indicating a worse performance. Additional figures are available in the Supp. Material.
	Both methods needed about 6 minutes to finish the estimation procedure. Attempts of the REML took longer than 9 hours without finishing the estimation and were aborted.

	\bibliographystyle{apalike}
	\bibliography{ref_V2}

\end{document}